\documentclass[journal=jacsat, manuscript=article]{achemso}
\SectionNumbersOn

\usepackage[utf8]{inputenc}
\usepackage{csquotes}
\usepackage{float}
\usepackage{microtype}
\usepackage{amsmath, amssymb, physics}
\usepackage{booktabs, bm, graphicx, xcolor}
\usepackage{subcaption}
\usepackage{url}
\usepackage[symbol]{footmisc}

\newcommand{\?}[1]{\ensuremath{\mathbf{#1}}}
\newcommand{\msp}[1]{\ensuremath{^\mathrm{#1}}}
\newcommand{\msb}[1]{\ensuremath{_\mathrm{#1}}}

\newcommand{\red}[1]{{\textcolor{black}{#1}}}   

\author{Sabine Käfer}
\affiliation{University of M\"unster,
Organisch-Chemisches Institut \\
and Center for Multiscale Theory and Computation,\\
Corrensstra{\ss}e 36, 48149 M\"unster, Germany}

\author{Niklas Niemeyer}
\affiliation{University of M\"unster,
Organisch-Chemisches Institut \\
and Center for Multiscale Theory and Computation,\\
Corrensstra{\ss}e 36, 48149 M\"unster, Germany}

\author{Johannes T\"{o}lle}
\affiliation{Division of Chemistry and Chemical Engineering, \\
California Institute of Technology, Pasadena, California 91125, USA}

\author{Johannes Neugebauer}
\affiliation{University of M\"unster,
Organisch-Chemisches Institut \\
and Center for Multiscale Theory and Computation,\\
Corrensstra{\ss}e 36, 48149 M\"unster, Germany}
\email{j.neugebauer@uni-muenster.de}

\title{Triplet Excitation-Energy Transfer Couplings 
from Subsystem Time-Dependent Density-Functional Theory}

\date{\today}

\begin{document}
~
\vfill
Date:~~December 12, 2023
\newpage
\begin{abstract}
We present an implementation of Triplet Excitation-Energy Transfer (TEET) couplings 
based on subsystem-based Time-Dependent Density-Functional Theory (sTDDFT). 
TEET couplings are systematically investigated by comparing ``exact'' 
and approximate variants of sTDDFT. 
We demonstrate that, while sTDDFT utilizing explicit approximate Non-Additive Kinetic 
Energy (NAKE) density functionals is well-suited for describing Singlet Excitation-Energy 
Transfer (SEET) processes, it is inadequate for characterizing TEET. 
However, we show that Projection-based Embedding (PbE)-based sTDDFT 
addresses the challenges faced by NAKE-sTDDFT and emerges as a promising method for 
accurately describing electronic couplings in TEET processes.

We also introduce the mixed PbE-/NAKE-embedding
procedure to investigate TEET effects in solvated pairs of chromophores. This 
approach offers a good balance between accuracy and efficiency, enabling 
comprehensive studies of TEET processes in complex environments.
\end{abstract}

\newpage

\section{Introduction}
Triplet Excitation-Energy Transfer (TEET) denotes the transfer of energy from one 
molecule in a triplet excited state to a singlet ground-state molecule under exchange 
of spin states. TEET plays an important role, for example, in
photosynthesis~\cite{Photosynthesis.1988,Photosynthesis.1993,Photosynthesis.2004},
where TEET reactions prevent the generation of undesired chlorophyll triplet states 
in order to protect photosynthetic organisms from harmful singlet oxygen,
as well as for organic photovoltaic~\cite{OPV.2005,OPV.2009} (OPV)
and organic light-emitting diode~\cite{OLED.2004} (OLED) devices.
Furthermore, TEET photocatalysis~\cite{Albini.1981,Glorius.2018,Zhou.2019,
Glorius.2020,Gilmour.2022} has gained significant 
importance in recent years in organic-synthetic photochemistry due to the challenges 
associated with directly preparing reactant molecules in triplet excited states.
With respect to compounds in which direct excitation is 
either hampered by a very low absorption coefficient or requires very harsh 
short-wavelength irradiation, TEET photocatalysis provides an alternative 
to a direct photoexcitation approach, which allows milder and more 
selective access to excited states. 
Visible-light-mediated energy-transfer-catalysis is, for example,
applied in the field of photoisomerization,
enabling the expansion of cyclization reactions~\cite{Smith.2017},
facilitating complex cascade reactions~\cite{Gilmour.2015,Gilmour.2016}, 
controlling systems with biological function~\cite{Weaver.2018}, 
and the rational design of powerful bond dissociation 
reactions~\cite{Glorius.EET.2018}. 
Additionally, it has been employed to accelerate or enable reaction-limiting 
steps in transition metal catalysis~\cite{MacMillan.2017}. 

The concept of TEET was first introduced by Dexter in 1953 and is arguably the 
best-known example of Dexter Energy Transfer processes~\cite{Dexter.1953}.
TEET involves the nonradiative transfer of excitation energy from an excited donor D* 
to an acceptor in its ground-state A 
\begin{equation}
    \mathrm{\textnormal{D*}}(\mathrm{T}_1) + \mathrm{A}(\mathrm{S}_0) 
    \rightarrow \mathrm{D}(\mathrm{S}_0) + \mathrm{\textnormal{A*}}(\mathrm{T}_1),
\label{EET: TEET}
\end{equation} 
leading to the formation of the excited state of the acceptor A*, 
while simultaneously regenerating the donor ground state D. TEET is often described 
as a simultaneous double-electron transfer between molecules without net changes in 
electron number or total spin states. We note, however, that more recent work has
pointed out the possibility of an alternative mechanism with step-wise electron 
transfer~\cite{Scholes.2021}.
In a simple two-level model, TEET can be described as follows:
we introduce the two diabatic states
\begin{align}
 \Xi_1 &= \ket{ \mathrm{\textnormal{D*}}(\mathrm{T}_1) \mathrm{A}(\mathrm{S}_0)}, \\
 \Xi_2 &= \ket{ \mathrm{\textnormal{A*}}(\mathrm{T}_1) \mathrm{D}(\mathrm{S}_0)},
\end{align}
whose coupling $V$ determines the TEET rate~\cite{Parson2023}, with $V$ given as
\begin{equation}\label{eq:two_level_V}
    V = \bra{\Xi_1} \hat{H} \ket{\Xi_2}.
\end{equation}
Only considering the respective two frontier orbitals of both the donor and acceptor 
within the diabatic states, TEET would entail the transfer of one electron form 
the Lowest Unoccupied Molecular Orbital (LUMO) referring to the ground-state occupation
of the excited triplet state of the donor $\phi_1^D$ to the LUMO of the ground-state acceptor 
$\phi_1^A$ with a simultaneous transfer of one electron with the opposite spin from 
the Highest Occupied Molecular Orbital (HOMO) of the acceptor $\phi_0^A$
to the HOMO of the donor $\phi_0^D$. 
One therefore finds for the normalized wavefunctions
\begin{align}
    \Xi_1 &= || \phi_0^{D,\alpha} \phi_1^{D,\alpha} \phi_0^{A,\alpha} \phi_0^{A,\beta}  ||, \label{eq:xi1}\\
    \Xi_2 &= || \phi_0^{D,\alpha} \phi_0^{D,\beta} \phi_0^{A,\alpha} \phi_1^{A,\alpha}  ||, \label{eq:xi2}
\end{align}
where $\alpha$ and $\beta$ denote spin functions. 
Inserting Eqs.~(\ref{eq:xi1})~and~(\ref{eq:xi2}) into Eq.~(\ref{eq:two_level_V}) 
yields Coulomb and exchange-type contributions to the electronic coupling
\begin{equation}
    V =\underbrace{\left( \phi_0^{D,\alpha} \phi_1^{D,\beta} | \phi_0^{A,\beta} \phi_1^{A,\alpha} \right)}_{=0}
    - \left( \phi_0^{D,\beta} \phi_0^{A,\beta} | \phi_1^{D,\alpha} \phi_1^{A,\alpha} \right),
\end{equation}
where the former vanishes for TEET. We note that formally electronic couplings only 
arise by introducing a diabatic basis, whose choice is inherently ambiguous.

While several quantum chemical methods have been successfully applied to 
Singlet Excitation-Energy Transfer (SEET) calculations,
TEET studies have not received as much attention.
The quantum-chemical description of TEET presents a higher level of complexity 
compared to SEET, primarily due to the challenges associated with the short-range 
distance dependence of TEET and the relatively small electronic couplings involved. 
The complexity of TEET arises from its dependence on the spatial overlap of the 
orbitals and charge distributions of the different molecules, leading to an exponential 
decrease in the TEET rate as the distance increases. 
Therefore, triplet energy transfer is limited to relatively short distances.
If donor and acceptor are connected via suitable bridging units, TEET may take place
via superexchange as described originally by McConnell~\cite{McConnell.1961} in 1961. 

Many methods developed for describing TEET couplings are based on wavefunction approaches, 
and are often formulated in a supersystem context. 
For example, Fleming and co-workers have presented two different approaches for 
determining TEET couplings, a direct coupling method as well as an 
energy-gap-based method within the Hartree--Fock (HF) theoretical 
framework~\cite{Fleming.2006}.
In the context of one-particle and two-particle coupling interactions, 
Beratan and co-workers have proposed a much simpler and more intuitive approach
to describe TEET couplings by predicting TEET interactions from Molecular Orbital (MO) 
overlaps with respect to direct and 
bridge-mediated donor--acceptor couplings~\cite{Beratan.2020}.

Subsystem Time-Dependent Density-Functional
Theory~\cite{Casida.Wesolowski.2004,Neugebauer.2007,Bockers.Neugebauer.2018,Neugebauer.2019}  
(subsystem TDDFT, sTDDFT) based on the Frozen-Density 
Embedding~\cite{Wesolowski.Warshel.1993} (FDE) approach
facilitates the description of excited states within a subsystem context
and can be used to investigate the excited-state properties of and excitation-energy
transfer couplings in complex systems, like e.g. the astaxanthin molecules in crustacyanin 
proteins~\cite{Astaxanthin.Neugebauer.2011},
the Fenna--Matthews--Olson (FMO) protein~\cite{LHC.FMO.Konig.2013},
and porphine nanotubes~\cite{Serestipy.Eschenbach.Niemeyer.2023}.
Based on its achievements in describing SEET~\cite{Toelle.2022}, it could be assumed that 
subsystem TDDFT might be well-suited for TEET as well. 
To the best of our knowledge, this route has not been pursued so far.
Therefore, the focus of this work will be on the possibility to describe TEET couplings 
with approximate~\cite{Neugebauer.2007} and ``exact''~\cite{Neugebauer.2019} variants of 
subsystem TDDFT within the coupled FDE framework.
In addition, we test a mixed procedure combining subsystem TDDFT making use of  
Projection-based Embedding~\cite{Manby.Miller.2012, Hegely.2016} (PbE) 
and approximate Non-Additive Kinetic Energy (NAKE) functionals  
in order to study TEET effects for solvated pairs of chromophores 
with respect to obtain a good balance between accuracy and efficiency.
This method will be referred to as \textit{mixed PbE-/NAKE-embedding} in the following.

This paper is structured as follows: 
First, the general theory of sTDDFT necessary to describe TEET within a subsystem 
context will be presented. Subsequently, the investigation of ``exact'' (PbE-sTDDFT) 
and approximate (NAKE-sTDDFT) subsystem TDDFT as well as the mixed  PbE-/NAKE-embedding,
will be discussed. Here, the focus is on the applicability of these methods as potential 
approaches for 
the description of TEET, starting with the analysis of several technical aspects.
In the second part of this work, PbE-sTDDFT and mixed PbE-/NAKE-embedding will be applied 
for investigating the impact of solvent effects on TEET couplings with respect to a 
perylene diimide (PDI) dimer system. In addition, the effect of bridge molecules on TEET 
couplings in bridge-mediated donor--acceptor systems will be investigated.

\section{Theory}
\subsection{Frozen-Density Embedding for Triplet States}\label{Triplet FDE}
Subsystem Time-Dependent Density-Functional Theory (TDDFT) is an extension of subsystem 
Density-Functional Theory (DFT)~\cite{Wesolowski.Warshel.1993,Wesolowski.Weber.1996,Neugebauer.2014} 
for describing excited states. Starting from Frozen-Density Embedding (FDE) as a practical 
realization of subsystem DFT, subsystem TDDFT is based on work by Casida and 
Weso{\l}owski~\cite{Casida.Wesolowski.2004} for embedded subsystems 
and was reformulated by one of us~\cite{Neugebauer.2007} to enable describing 
delocalized excitation processes in interacting chromophore 
aggregates~\cite{Konig.Schlüter.Neugebauer.2013}.
In the unrestricted generalization of subsystem TDDFT~\cite{Bockers.Neugebauer.2018},
excitation energies $\omega$ are determined as solutions of the non-Hermitian eigenvalue problem 
\begin{gather}
    \begin{pmatrix} \mathbf{A} & \mathbf{B} \\ \mathbf{B} & \mathbf{A} \end{pmatrix}
    \begin{pmatrix} \mathbf{X} \\ \mathbf{Y} \end{pmatrix}
    =
    \omega \begin{pmatrix} \mathbf{-1} & \mathbf{0} \\ \mathbf{0} & \mathbf{1} \end{pmatrix}
    \begin{pmatrix} \mathbf{X} \\ \mathbf{Y} \end{pmatrix},
\label{sTDDFT: sTDDFTequation_2}
\end{gather}
whose regular TDDFT analog is commonly referred to as Casida’s equation~\cite{Casida.1995}.
In the subsystem context, the elements of the response matrices $\?A$ and $\?B$ are defined as
\begin{align}
    A_{(ia\sigma)_I,(jb\tau)_J} &= \delta_{IJ} \delta_{ij} \delta_{ab} \delta_{\sigma\tau}
    (\epsilon_{a\sigma}^I - \epsilon_{i\sigma}^I) 
    + K_{(ia\sigma)_I,(jb\tau)_J}, 
\label{sTDDFT: MatrixA} \\
    B_{(ia\sigma)_I,(jb\tau)_J} &= K_{(ia\sigma)_I,(bj\tau)_J}, 
\label{sTDDFT: MatrixB}
\end{align}
with elements of the coupling matrix $\?K$ (see below).
Here and in the following, $i,j$ and $a,b$ denote occupied and virtual reference 
orbitals, respectively. Subscripts $I,J$ label the subsystems involved and spin functions
are denoted by $\sigma$ and $\tau \in \alpha,\beta$.
The eigenvectors $(\?X\>\>\?Y)^T$ gather elements of the off-diagonal blocks of the transition density matrix $\delta D$
for each excitation
\begin{align}
    X_{(ia\sigma)_I} &= \delta D_{(ia\sigma)_I},  \\
    Y_{(ia\sigma)_I} &= \delta D_{(ai\sigma)_I}.
\label{sTDDFT:TrDens}
\end{align}
Setting up the eigenvalue equation in Eq.~(\ref{sTDDFT: sTDDFTequation_2}) starting from
a restricted closed-shell subsystem DFT reference leads to pairs of degenerate orbital transitions (and associated orbital-energy differences) for each spin channel and each subsystem. For the spin blocks of the coupling matrix one finds that
\begin{align}
    K_{\alpha\alpha} = K_{\beta\beta}, \\
    K_{\alpha\beta} = K_{\beta\alpha}.
\end{align}
As a result, the eigenvalue problem in Eq.~(\ref{sTDDFT: sTDDFTequation_2}) decouples into two separate eigenvalue equations of identical structure with symmetric $(+)$ and antisymmetric $(-)$ combinations of the coupling matrix
\begin{align}
    K^{(\pm)} = K^{(\pm)}_{\alpha\alpha} \pm K^{(\pm)}_{\alpha\beta}.
\end{align}
The singlet and triplet eigenvalue problems can therefore be derived as simple unitary transformations of the unrestricted response problem with eigenstates
\begin{align}
    \delta D^{(\pm)}_{(pq)_I} &= \frac{1}{\sqrt{2}} \left(\delta D_{(pq\alpha)_I} \pm \delta D_{(pq\beta)_I}\right), \quad pq\in ia, ai,
\end{align}
directly enabling to discriminate between singlet (which are spin-symmetry-conserving and correspond to the $(+)$ combination) and triplet excitations (which are spin-symmetry breaking and correspond to the $(-)$ combination)~\cite{Ahlrichs.1996}.

While the orbital-energy differences on the diagonal of the $\?A$ matrix are retained for both the singlet and
triplet eigenvalue equation, the coupling matrix requires further consideration. We refer the 
reader to Refs.~\citenum{Toelle.2019b} and \citenum{Niemeyer.2020} for a more detailed discussion of 
the singlet coupling matrix of subsystem TDDFT.
Using Mulliken's notation for two-electron integrals, 
elements of the triplet coupling matrix $\?{K^{(-)}}$ can be written as
\begin{multline}
    K_{(ia)_I, (jb)_J}^{(-)} = \delta_{IJ}\bigl[\left(\phi_i^I\phi_a^I\left| f\msb{xc}^{I,\alpha\alpha} 
    - f\msb{xc}^{I,\alpha\beta} \right|\phi_j^J\phi_b^J\right)
    - c\msb{HF}^I\left(\phi_i^I\phi_j^J|\phi_a^I\phi_b^J\right)\bigr] \\
    + \left(\phi_i^I\phi_a^I\left| f\msb{xc}\msp{nadd,\alpha\alpha} 
    - f\msb{xc}\msp{nadd,\alpha\beta} \right|\phi_j^J\phi_b^J\right)
    + \left(\phi_i^I\phi_a^I\left| f\msb{kin}\msp{nadd,\alpha\alpha} 
    - f\msb{kin}\msp{nadd,\alpha\beta} \right|\phi_j^J\phi_b^J\right),
\label{FDE: TripletCouplings}
\end{multline}
with the exchange--correlation (XC) energy kernel of subsystem $I$
\begin{equation}
    f^{I,\sigma\tau}_\mathrm{xc}(\mathbf{r},\mathbf{r'}) 
    = \left.\frac{\delta^2 E_{\mathrm{xc}}[\rho]}{\delta\rho^{\sigma}(\mathbf{r})
    \delta\rho^{\tau}(\mathbf{r'})} \right|_{\rho=\rho_I},
\label{sTDDFT: XCKernel}
\end{equation}
where $c^I\msb{HF}$ is the amount of Hartree--Fock exchange in the functional approximation used for
evaluating $f^{I}_\mathrm{xc}$. Likewise, one finds for the non-additive kinetic and XC energy kernels, $f\msb{kin}\msp{nadd}$ and $f\msb{xc}\msp{nadd}$ with energy functionals $F=T_s,E\msb{xc}$, that
\begin{equation}
    f^{\mathrm{nadd},\sigma\tau}(\mathbf{r},\mathbf{r'}) 
    = \left.\frac{\delta^2 F[\rho]}{\delta\rho^\sigma(\mathbf{r})
    \delta\rho^\tau(\mathbf{r'})} \right|_{\rho=\rho_{\mathrm{tot}}} 
    - \delta_{IJ} \left.\frac{\delta^2 F[\rho]}{\delta\rho^\sigma(\mathbf{r})
    \delta\rho^\tau(\mathbf{r'})} \right|_{\rho=\rho_I}.
\label{sTDDFT: NAKE/XCKernel}
\end{equation}
Following the adiabatic approximation, the subsystem densities $\rho_I$ and the supersystem density $\rho\msb{tot}$ are evaluated with the corresponding electronic ground-state solutions.
For the sake of completeness, the full triplet subsystem TDDFT eigenvalue equation therefore reads
\begin{gather}
    \begin{pmatrix} \mathbf{\bar A} & \mathbf{\bar B} \\ \mathbf{\bar B} & \mathbf{\bar A} \end{pmatrix}
    \begin{pmatrix} \mathbf{\bar X} \\ \mathbf{\bar Y} \end{pmatrix}
    =
    \bar \omega \begin{pmatrix} \mathbf{-1} & \mathbf{0} \\ \mathbf{0} & \mathbf{1} \end{pmatrix}
    \begin{pmatrix} \mathbf{\bar X} \\ \mathbf{\bar Y} \end{pmatrix},
\label{sTDDFT: sTDDFTequation_Triplet}
\end{gather}
with
\begin{align}
    \bar A_{(ia)_I,(jb)_J} &= \delta_{IJ} \delta_{ij} \delta_{ab} 
    (\epsilon_{a}^I - \epsilon_{i}^I) 
    + K_{(ia)_I,(jb)_J}^{(-)}, \\
    \bar B_{(ia)_I,(jb)_J} &= K_{(ia)_I,(bj)_J}^{(-)}.
\end{align}
The key differences to the singlet eigenvalue problem are (i) the Coulomb coupling matrix elements vanish and (ii) different functional derivatives need to be evaluated for each component of the kernel. The main challenge for the extension of an existing singlet or unrestricted subsystem TDDFT code for computing triplet excitations from restricted references is to evaluate the appropriate functional derivatives for the numerical integration of the XC as well as non-additive XC and kinetic kernels. For a discussion of the corresponding terms in regular TDDFT, see Ref.~\citenum{Ahlrichs.1996}. On the contrary, however, Hartree--Fock exchange contributions to the triplet coupling matrix are the same as in the singlet case.
Further, due to the conserved structure of the eigenvalue problem, identical (iterative) solution strategies as used for Eq.~(\ref{sTDDFT: sTDDFTequation_2}) can be applied for Eq.~(\ref{sTDDFT: sTDDFTequation_Triplet}).
This work focuses on the Tamm--Dancoff Approximation (TDA)~\cite{Hirata.Head-Gordon.1999} of subsystem TDDFT~\cite{Konig.Schlüter.Neugebauer.2013}.
Neglecting the $\?B$ matrix in the eigenvalue problem, the subsystem TDDFT eigenvalue problem 
turns into a Hermitian eigenvalue problem of reduced dimension
\begin{equation}
    \mathbf{\bar A} \mathbf{\bar X} = \bar\omega \mathbf{\bar X},
\label{sTDDFT: TDA-sTDDFT}  
\end{equation}
allowing for a more straightforward diagonalization process. Another important
implication of the TDA in the subsystem TDDFT context is that electronic couplings 
can be calculated directly~\cite{Konig.Schlüter.Neugebauer.2013}, as outlined below.

Subsystem TDDFT in practice is typically performed in a two-step procedure as first outlined in Ref.~\citenum{Neugebauer.2007}.
In the first step, referred to as uncoupled Frozen-Density Embedding (FDEu), some of the lowest-lying
excitations of each subsystem are determined separately. This is done by (partially) diagonalizing
the intra-subsystem blocks of the response matrices appearing in Eqs.~(\ref{sTDDFT: sTDDFTequation_2}) 
or (\ref{sTDDFT: sTDDFTequation_Triplet}), yielding a set of FDEu excitation energies and
transition densities.
In the subsequent step, known as coupled Frozen-Density Embedding (FDEc), 
the obtained FDEu transition densities are first used to construct 
a transformation matrix $\?U$, which is then employed to transform the full subsystem
(TDA-)TDDFT response problem into the subspace spanned by the FDEu excitation vectors
\begin{align}
    \mathbf{U^T} \mathbf{\bar A} \mathbf{U} \mathbf{U^T} \mathbf{\bar X} 
    &= \omega^{\mathrm{FDEc}} \mathbf{U^T} \mathbf{\bar X}, \label{FDEc: FDEcProblem} \\
    \tilde{{\bar{\?A}}} \tilde{{\bar{\?X}}} &= \omega^{\mathrm{FDEc}} \tilde{{\bar{\?X}}},
\label{FDEc: FDEcProblem_2}  
\end{align}
introducing the FDEc subspace matrix $\tilde{{\bar{\?A}}}$ with elements
\begin{equation}
    \Tilde{\mathbf{\bar A}}_{(m)_I(n)_J}
    = \sum_{(ia)_I} \sum_{(jb)_J} \bar X_{(ia)_I}^m \bar A_{(ia)_I, (jb)_J} \bar X_{(jb)_J}^n,
\label{FDEc: Hamilton-likeMatrix}
\end{equation}
where subscripts $(m)_I$ and $(n)_J$ denote FDEu excitations $m$ and $n$ of subsystem
$I$ and $J$, respectively.
The block structure of the subsystem TDA-TDDFT eigenvalue problem implies that
Eq.~(\ref{FDEc: Hamilton-likeMatrix}) can be conveniently rewritten as
\begin{equation}
    \Tilde{\mathbf{\bar A}}_{(m)_I(n)_J} = \delta_{IJ} \delta_{mn} \omega_m^I 
        + (1-\delta_{IJ}) V_{(m)_I(n)_J},
\label{FDEc: Hamilton-likeMatrix_2}  
\end{equation}
such that off-diagonal blocks contain the electronic couplings
$V_{(m)_I(n)_J}$ and the diagonal comprises the uncoupled excitation energies $\omega_m^I$.
Following an exciton picture, the $\tilde{{\bar{\?A}}}$ matrix can thus be 
interpreted as a triplet Hamilton-like matrix.
To avoid any ambiguity, we semantically distinguish between (i) 
the electronic couplings $V$ extracted from the off-diagonal 
blocks of the FDEc subspace matrix (which are interpreted as Triplet Excitation-Energy
Transfer (TEET) couplings in this work)
and (ii) elements of the triplet coupling matrix $\?{K^{(-)}}$.

\subsection{Projection-based Embedding}
Similar to FDE in the original sense~\cite{Wesolowski.Warshel.1993,Wesolowski.Weber.1996},
projection-based embedding (PbE) partitions the supersystem density into subsystem densities.
To avoid issues associated with the non-additivity of the subsystem kinetic energies in FDE,
projection-based embedding enforces inter-subsystem or external-orthogonality (EO) of
the subsystem orbitals by virtue of a non-local projector $\hat{P}_B$,
\begin{equation}
    \hat{P}_B = \sum_j^{n_B} \ket{\phi_j^B} \bra{\phi_j^B},
\label{PbE: Projector}  
\end{equation}
for a  system consisting of an active system $A$ embedded in an 
environmental subsystem $B$ with $n_B$ occupied environment orbitals $\phi_j^B$.
Projection-based embedding in practice involves adding a particular projection operator $\hat{O}$ to
the supersystem Fock operator $\hat{F}$ to obtain the active subsystem's orbitals $\phi_i^A$,
\begin{equation}
    (\hat{F} + \hat{O}) \ket{\phi_i^A} = \epsilon_i^A \ket{\phi_i^A}.
\end{equation}
Several projection operators have been presented in the literature. This work focuses
on the levelshift operator~\cite{Manby.Miller.2012} 
\begin{equation}
   \hat{O}\msp{level.} = \lim \limits_{\mu \to \infty} \mu \hat{P}_B,
\label{PbE: LevelshiftOperator}  
\end{equation}
with a levelshift parameter $\mu$, as well as on the Huzinaga operator~\cite{Huzinaga.1971}
\begin{equation}
    \hat{O}\msp{Huz.} = - [\hat{F}, \hat{P}^B]_+.
\label{PbE: HuzinagaOperator} 
\end{equation}
The usage of projection operators formally replaces NAKE kernel contributions of the coupling
matrix $\?K$ with density matrix ($\?D$) derivatives of the EO potential in matrix representation 
$\?V^\mathrm{EO}$. These were originally derived in Ref.~\citenum{Chulhai.2016} for the levelshift operator as well as corrected and further discussed in Refs.~\citenum{Neugebauer.2019} and \citenum{Toelle.2019b}.
We find for EO kernel contributions of the levelshift operator to the triplet coupling matrix, 
\begin{align}
    K^{(-),\mathrm{EO, level.}}_{(ia)_I,(jb)_J} 
        &= \frac{\partial V_{(ia\alpha)_I}^{\mathrm{EO, level.}}}{\partial D_{(jb\alpha)_J}} -
           \frac{\partial V_{(ia\alpha)_I}^{\mathrm{EO, level.}}}{\partial D_{(jb\beta)_J}} \\
        &= \mu\langle\phi_{i\alpha}^I|\phi_{j\alpha}^J\rangle \langle\phi_{b\alpha}^J|\phi_{a\alpha}^I\rangle -
        \mu\langle\phi_{i\alpha}^I|\phi_{j\beta}^J\rangle \langle\phi_{b\beta}^J|\phi_{a\alpha}^I\rangle \\
        &= \mu \langle\phi_{i}^I|\phi_{j}^J\rangle \langle\phi_{b}^J|\phi_{a}^I\rangle, 
\label{Exact sTDDFT: Level_MatrixA}
\end{align}
where the second summand trivially vanishes due to spin-function orthogonality rendering the resulting
expression identical to the singlet case. Similar arguments can be given for the
Huzinaga EO kernel~\cite{Toelle.2019b},
\begin{align}
     K^{(-),\mathrm{EO,Huz.}}_{(ia)_I,(jb)_J} 
        &= \frac{\partial V^\mathrm{EO,Huz.}_{(ia\alpha)_I}}{\partial D_{(jb\alpha)_J}} - \frac{\partial V^\mathrm{EO,Huz.}_{(ia\alpha)_I}}{\partial D_{(jb\beta)_J}}  \\
        &= - \langle\phi_{i}^I|\hat F|\phi_{j}^J\rangle \langle\phi_{b}^J|\phi_{a}^I\rangle.
\label{Exact sTDDFT: Huz_MatrixA}
\end{align}
Collecting terms, elements of the triplet coupling matrix in subsystem TDDFT with projection-based
embedding read
\begin{multline}
    K^{(-)}_{(ia)_I, (jb)_J} 
    = \delta_{IJ}\bigl[\left(\phi_i^I\phi_a^I\left| f\msb{xc}^{I,\alpha\alpha} 
    - f\msb{xc}^{I,\alpha\beta} \right|\phi_j^J\phi_b^J\right)
    - c\msb{HF}^I\left(\phi_i^I\phi_j^J|\phi_a^I\phi_b^J\right)
    + c\msb{HF}\msp{nadd}\left(\phi_i^I\phi_j^J|\phi_a^I\phi_b^J\right)\bigr] \\
    + \left(\phi_i^I\phi_a^I\left| f\msb{xc}\msp{nadd,\alpha\alpha} 
    - f\msb{xc}\msp{nadd,\alpha\beta} \right|\phi_j^J\phi_b^J\right) - c\msb{HF}\msp{nadd}\left(\phi_i^I\phi_j^J|\phi_a^I\phi_b^J\right) + K^{(-),\mathrm{EO}}_{(ia)_I,(jb)_J},
\label{PbE: TripletCoupling}
\end{multline}
where inter-subsystem orbital orthogonality allows for inter-subsystem Hartree--Fock 
exchange via the functional approximation used for evaluating the non-additive XC kernel
(which can, in principle, be chosen independently of the intra-subsystem XC approximation). 
The corresponding amount is labeled $c\msb{HF}\msp{nadd}$.

This work further employs the Fermi-shifted Huzinaga potential~\cite{Chulhai.2018}, for which, 
to the best of our knowledge, no EO kernel has been presented in the literature yet.
A brief derivation is given in the appendix of this article.

\subsection{Mixed PbE-/NAKE-Embedding} \label{HybridEmbedding}
\begin{figure}[!t]
    \centering
    \includegraphics[width=0.5\textwidth]{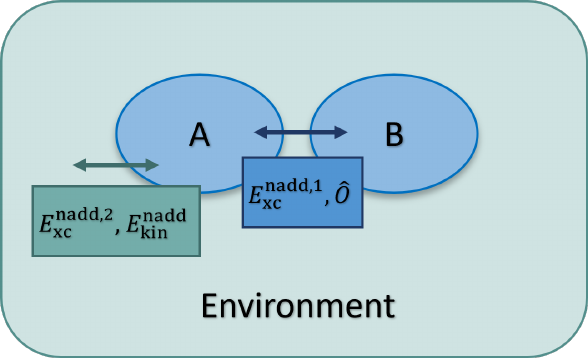}
\caption{
Illustration of mixed PbE-/NAKE-embedding with 
subsystem $A$ and $B$ being described using PbE-sTDDFT
with inter-subsystem exchange--correlation functional $E\msb{xc}\msp{nadd,1}$ 
and projection operator $\hat{O}$
and the environment using NAKE-sTDDFT
with $E\msb{xc}\msp{nadd,2}$ 
and non-additive kinetic energy functional $E\msb{kin}\msp{nadd}$.
}
\label{fig1}
\end{figure}
In order to enable a more efficient description of complex systems, we introduce a 
combined embedding approach for investigating environmental effects on sub-parts of the 
considered system. The idea of this \textit{mixed PbE-/NAKE-embedding}~\cite{Tolle.PhD.2021} 
procedure, schematically depicted in Fig.~\ref{fig1}, is the partitioning of 
the electronic density into two subspaces belonging to $N\msb{PbE}$ subsystems described 
using PbE-sTDDFT with density $\rho\msb{PbE}$ (exact embedding), and $N\msb{NAKE}$ subsystems 
described by NAKE-sTDDFT with density $\rho\msb{NAKE}$ (approximate embedding),
\begin{align}
    \rho\msb{tot}(r) &= \rho\msb{PbE}(\mathbf{r}) + \rho\msb{NAKE}(\mathbf{r}) \\
        &= \sum_I^{N\msb{PbE}} \rho_I(\mathbf{r}) + \sum_J^{N\msb{NAKE}} \rho_J(\mathbf{r}).  
\end{align}
This partitioning is illustrated in Fig.~\ref{fig1}, where 
$\rho\msb{PbE}$ corresponds to the density of the subsystems A and B, 
while $\rho\msb{NAKE}$ corresponds to the environment density.
Within each subspace, the densities are obtained from an arbitrary number 
of assigned subsystems.
The system's total energy is expressed as a combination of Kohn--Sham energies, 
interaction energies, and two different non-additive exchange--correlation contributions
\begin{align}
    E = &\sum_I^{N\msb{PbE}} E\msp{KS}[\rho_I] + \sum_J^{N\msb{NAKE}} E\msp{KS}[\rho_J] 
        + J[\{\rho_K\}] + V\msb{nuc,el}[\{\rho_K\}]  \nonumber \\
        &+ E\msb{xc}\msp{nadd,1}[\rho\msb{PbE},\{\rho_I\}] 
        + E\msb{xc}\msp{nadd,2}[\rho\msb{tot},\rho\msb{PbE},\{\rho_J\}] \\
        &+ T_s\msp{nadd}[\rho\msb{tot},\rho\msb{PbE},\{\rho_J\}], \nonumber  
\end{align}
one associated with the PbE region $E\msb{xc}\msp{nadd,1}$ 
and the other with the remaining part of the system $E\msb{xc}\msp{nadd,2}$.
For the non-additive part associated with the PbE region, the non-additive
exchange-correlation contribution is given by
\begin{equation}
    E\msb{xc}\msp{nadd,1}[\rho\msb{PbE},\{\rho_I\}]
        = E\msb{xc}[\rho\msb{PbE}] - \sum_I^{N\msb{PbE}} E\msb{xc}[\rho_I],
\end{equation}
and for the remaining part of the system, the non-additive XC functional 
contribution can be determined as
\begin{equation}
    E\msb{xc}\msp{nadd,2}[\rho\msb{tot},\rho\msb{PbE},\{\rho_J\}]
        = E\msb{xc}[\rho\msb{tot}] - \sum_J^{N\msb{NAKE}} E\msb{xc}[\rho_J] 
        - E\msb{xc}[\rho\msb{PbE}].
\end{equation}
Note that in principle different approximate XC functionals can be used for the 
evaluation of these contributions.
The non-additive kinetic energy is calculated similarly,
\begin{equation}
    T_s\msp{nadd}[\rho\msb{tot},\rho\msb{PbE},\{\rho_J\}]
        = T_s[\rho\msb{tot}] - \sum_J^{N\msb{NAKE}} T_s[\rho_J] 
        - T_s[\rho\msb{PbE}].
\end{equation}
The Kohn--Sham equations with constrained electron density (KSCED) 
within the mixed PbE-/NAKE-embedding approach are given as
\begin{equation}
    \left(
    - \frac{\nabla^2}{2} + \hat{v}\msb{eff}^K + \hat{v}\msb{emb}^K
    \right) 
    \phi_i^K = \epsilon_i^K \phi_i^K,
\end{equation}
where $\hat{v}\msb{eff}^K$ is the Kohn--Sham effective potential evaluated for subsystem \textit{K}.
The embedding potential $\hat{v}\msb{emb}^K$ for a subsystem \textit{K} in the NAKE region
can be expressed as
\begin{align}
    \hat{v}\msb{emb}^K[\rho_K,\rho\msb{PbE},\rho\msb{tot}] (\mathbf{r})
    = &\sum^{N\msb{NAKE}+N\msb{PbE}}_{L \neq K} \left[ \int \frac{\rho_L(\mathbf{r'})}{|\mathbf{r'} - \mathbf{r}|} \mathrm{d}\mathbf{r'}
    - \sum_{\alpha \in L} \frac{Z_{\alpha}}{|\mathbf{r} - \mathbf{R}_{\alpha}|} \right] \nonumber \\   
    &+ \frac{\delta E\msb{xc}[\rho\msb{tot}]}{\delta\rho\msb{tot}(\mathbf{r})} 
    - \frac{\delta E\msb{xc}[\rho_K]}{\delta\rho_K(\mathbf{r})} \\
    &+ \frac{\delta T_s[\rho\msb{tot}]}{\delta\rho\msb{tot}(\mathbf{r})}
    - \frac{\delta T_s[\rho_K]}{\delta\rho_K(\mathbf{r})}. \nonumber
\end{align}
For a subsystem \textit{K} in the PbE region, $\hat{v}\msb{emb}^K$ becomes a non-local operator
through the projection operator $\hat{O}$ [see Eqs.~\eqref{PbE: LevelshiftOperator} 
and \eqref{PbE: HuzinagaOperator}] and reads
\begin{align}
    \hat{v}\msb{emb}^K
    = &\sum^{N\msb{NAKE}+N\msb{PbE}}_{L \neq K} \left[ \int \frac{\rho_L(\mathbf{r'})}{|\mathbf{r} - \mathbf{r'}|} \mathrm{d}\mathbf{r'}
    - \sum_{\alpha \in L} \frac{Z_{\alpha}}{|\mathbf{r} - \mathbf{R}_{\alpha}|} \right] \nonumber \\   
    &+ \frac{\delta E\msb{xc}[\rho\msb{tot}]}{\delta\rho\msb{tot}(\mathbf{r})}
    - \frac{\delta E\msb{xc}[\rho\msb{PbE}]}{\delta\rho\msb{PbE}(\mathbf{r})} 
    + \frac{\delta \Tilde{E}\msb{xc}[\rho\msb{PbE}]}{\delta\rho\msb{PbE}(\mathbf{r})}
    - \frac{\delta \Tilde{E}\msb{xc}[\rho_K]}{\delta\rho_K(\mathbf{r})} \\
    &+ \frac{\delta T_s[\rho\msb{tot}]}{\delta\rho\msb{tot}(\mathbf{r})}
    - \frac{\delta T_s[\rho\msb{PbE}]}{\delta\rho\msb{PbE}(\mathbf{r})} + \hat{O} \nonumber,
\end{align}
where the tilde refers to the XC potential contributions arising from 
the non-additive XC energy associated with the PbE region.

In the sTDDFT context, the mixed PbE-/NAKE-embedding coupling matrix is likewise
obtained from derivatives of matrix elements of the electronic potential of subsystem $I$
\begin{equation}
    \hat{v}\msb{el}^I =  \hat{v}\msb{eff}^I +  \hat{v}\msb{emb}^I
\end{equation}
with respect to elements of the density matrix $\?D$ of subsystem $K$ 
\begin{equation}
  K_{(ia)_I,(jb)_K} = \frac{\partial v_{\mathrm{el},(ia)}^I}{\partial D_{(jb)_K}},
\end{equation}
and is defined as (in the non-spin-resolved case) 
\begin{multline}
    K_{(ia)_I,(jb)_K} = 
    \left(\phi_i^I\phi_a^I|\phi_j^K\phi_b^K\right) + \left(\phi_i^I\phi_a^I\left|f\msb{xc}\msp{tot} + f\msb{kin}\msp{tot} - f\msb{xc}\msp{PbE} - f\msb{kin}\msp{PbE} + \Tilde{f}\msb{xc}\msp{PbE}  \right| \phi_j^K\phi_b^K\right) \\
     + \delta_{IK}\left[\left(\phi_i^I\phi_a^I\left|f\msb{xc}^{\mathrm{intra},I} - \Tilde{f}\msb{xc}^{\mathrm{PbE},I}\right| \phi_j^K\phi_b^K\right) \right] + K_{(ia)_I,(jb)_K}\msp{EO}
\label{Hybrid:Kernel1} 
\end{multline}
for $I$ and $K$ being both part of the PbE region. With $F=T_s,E\msb{xc}$, the various kinetic and exchange--correlation kernel contributions in Eq.~(\ref{Hybrid:Kernel1}) follow the definitions
\begin{align}
f^{\mathrm{tot}}(\mathbf{r},\mathbf{r'}) 
    &= \left.\frac{\delta^2 F[\rho]}{\delta\rho(\mathbf{r})
    \delta\rho(\mathbf{r'})} \right|_{\rho=\rho\msb{tot}}, \\
f^{\mathrm{PbE}}(\mathbf{r},\mathbf{r'}) 
    &= \left.\frac{\delta^2 F[\rho]}{\delta\rho(\mathbf{r})
    \delta\rho(\mathbf{r'})} \right|_{\rho=\rho\msb{PbE}}, \\
\tilde f^{\mathrm{PbE}}_\mathrm{xc}(\mathbf{r},\mathbf{r'}) 
    &= \left.\frac{\delta^2 \tilde E_\mathrm{xc}[\rho]}{\delta\rho(\mathbf{r})
    \delta\rho(\mathbf{r'})} \right|_{\rho=\rho\msb{PbE}}, \\
 f^{\mathrm{intra},I}_\mathrm{xc}(\mathbf{r},\mathbf{r'}) 
    &= \left.\frac{\delta^2 E_\mathrm{xc}[\rho]}{\delta\rho(\mathbf{r})
    \delta\rho(\mathbf{r'})} \right|_{\rho=\rho_I}, \\
\tilde f^{\mathrm{PbE},I}_\mathrm{xc}(\mathbf{r},\mathbf{r'}) 
    &= \left.\frac{\delta^2 \tilde E_\mathrm{xc}[\rho]}{\delta\rho(\mathbf{r})
    \delta\rho(\mathbf{r'})} \right|_{\rho=\rho_I}.
\end{align}
In the case of $I$ being part of the PbE region and $K$ of the NAKE region,
the kernel is given as 
\begin{equation}
    K_{(ia)_I,(jb)_K} = \left(\phi_i^I\phi_a^I|\phi_j^K\phi_b^K\right) + \left(\phi_i^I\phi_a^I\left|f\msb{xc}\msp{tot} + f\msb{kin}\msp{tot} \right| \phi_j^K\phi_b^K\right),
\label{Hybrid:Kernel2}
\end{equation}
and for $I$ and $K$ being both part of the NAKE region, it can be expressed as
\begin{equation}
    K_{(ia)_I,(jb)_K} = \left(\phi_i^I\phi_a^I|\phi_j^K\phi_b^K\right) + \left(\phi_i^I\phi_a^I\left|f\msb{xc}\msp{nadd} + f\msb{kin}\msp{nadd} \right| \phi_j^K\phi_b^K\right)
    + \delta_{IK}\left(\phi_i^I\phi_a^I\left|f\msb{xc}^{\mathrm{intra},I}\right| \phi_j^K\phi_b^K\right),
\label{Hybrid:Kernel3}
\end{equation}
where the non-additive contributions follow directly from Eq.~(\ref{sTDDFT: NAKE/XCKernel}) (without the explicit inclusion of HF exchange in any exchange--correlation functional approximation for brevity).
Note that Eq.~\eqref{Hybrid:Kernel2} only becomes relevant for FDEc-TDDFT calculations,
while Eqs.~\eqref{Hybrid:Kernel1} and \eqref{Hybrid:Kernel3} may also be needed in 
FDEu-TDDFT calculations for the case of identical subsystem labels $I, K$.
This allows to describe the subspace of interest using ``exact'' PbE embedding 
while describing the remaining subsystems associated to the environment 
with approximate embedding. Therefore, environmental effects like the effect of
an explicit solvent on electronic couplings can be studied more efficiently
without compromising accuracy for the coupled chromophores. 

For mixed PbE-/NAKE-embedding, the following notation will be used in this article: 
Intra-XC/[Inter1-XC, Inter2-XC]/[$\hat{O}$, NAKE func.].
Here, Intra-XC refers to the XC functional for intra-subsystem interactions, 
Inter1-XC denotes the inter-subsystem XC functional of the two active subsystems 
and Inter2-XC is used for the remaining inter-subsystem interactions. 
Additionally, the projection operator $\hat{O}$ for the exact embedding approach is 
given as well as the NAKE functional, denoted as NAKE func.,
used in the approximate embedding approach.

\section{Computational Details}
All technical test calculations in the following were performed using the helium dimer
and fluoroethylene dimer as test systems. The geometry of the latter
was optimized with \textsc{Turbomole 7.5.1}~\cite{TURBOMOLE} 
using PBE0~\cite{PBE} as the exchange--correlation functional 
and a def2-TZVP~\cite{Def2_Basis} basis set.
The structures of the perylene diimide (PDI) dimer in water were taken from 
Ref.~\citenum{PDI.Struc.Curutchet.2009}, while those for the PDI dimer in other  
solvents were obtained from Ref.~\citenum{PDI.Curutchet.JPC.2012}.
All calculations were performed with a slightly modified version of
the \textsc{Serenity} program (1.5.2)~\cite{serenity,serenity2,serenity.zenodo}. 
Unless stated otherwise, all supersystem and subsystem TDDFT calculations 
were performed with this version of \textsc{Serenity}
using the PBE0 XC functional and the def2-SVP~\cite{Def2_Basis} basis set. 
All self-consistent field (SCF) procedures were stopped
as soon as two of the following convergence criteria were fulfilled: total energy threshold
of $5\cdot 10^{-8}$~E$_\mathrm{h}$, root-mean-square deviation of the density matrix threshold of $1\cdot 10^{-8}$~a.u., as
well as a threshold of $5\cdot 10^{-7}$~a.u. for the commutator of the Fock and density matrix.
The intrinsic-bond orbital localization method\cite{Knizia.2013} is used in
projection-based embedding calculations.
Moreover, the Tamm--Dancoff approximation (TDA) was used in each calculation and
the six lowest-lying excitation energies for each subsystem 
and the twelve lowest-lying excitation energies for each supersystem were determined.
The eigenvectors of the iterative solution procedure of the TDA problem were converged to maximum residual norms of $10^{-5}$.
Additionally, for embedding calculations, PW91~\cite{PW91} was used 
as the non-additive XC functional and PW91k~\cite{PW91k} as the 
non-additive kinetic-energy functional in the case of NAKE-sTDDFT calculations. 
In case of PbE-sTDDFT, PBE0 was chosen as the non-additive XC functional  
in combination with the levelshift projection operator if not mentioned otherwise.
All mixed PbE-/NAKE-embedding calculations were performed using the following 
settings: [PBE0, PBE]/[PBE0, PBE]/[level., PW91K], def2-SVP.
Note that, unless stated otherwise, the obtained results were determined for 
excitations dominated by HOMO($\pi$) $\rightarrow$ LUMO($\pi$*) transitions
and their corresponding electronic couplings.

\section{Results and Discussion}
\subsection{Subsystem TDDFT for TEET Couplings}
\subsubsection{Helium Dimer}

\begin{figure}[!t]
    \centering
    \includegraphics[width=\textwidth]{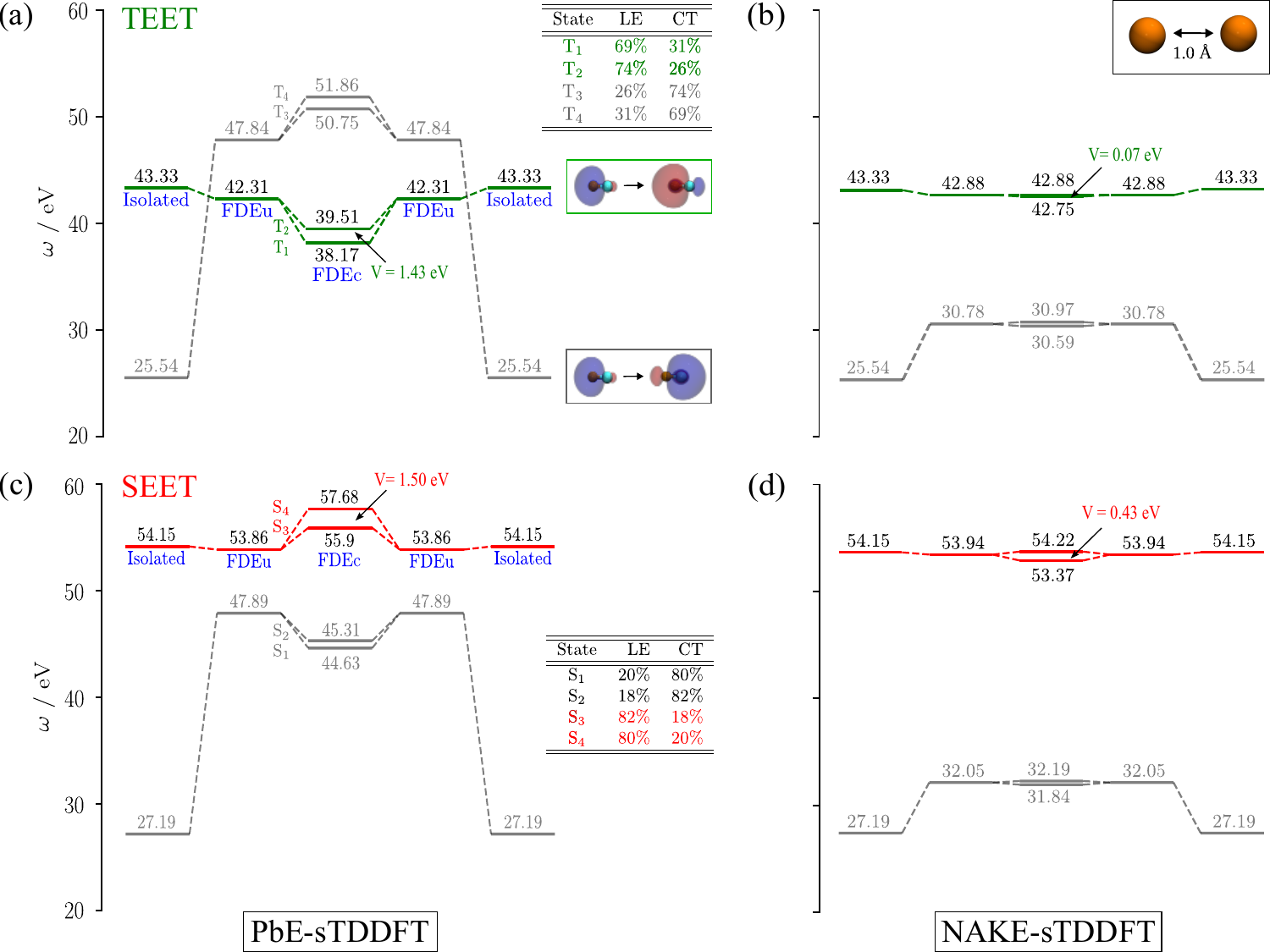}
\caption{
Excitation energies $\omega$ and electronic couplings $V$ (in eV), represented 
by colored values, of the triplet (green) and singlet (red) HOMO $\rightarrow$ LUMO 
local excitation (LE) of the helium dimer at 1.0~\r{A} separation
calculated using TDDFT [PBE0, 6-31G] for isolated systems, 
as well as uncoupled (FDEu), and coupled (FDEc). 
(a) and (c), PbE-sTDDFT [PbE: PBE0/PBE0/Fermi-shifted Huzinaga, 6-31G];
(b) and (d), NAKE-sTDDFT [NAKE: PBE0/PW91/PW91K, 6-31G].
Gray lines correspond to the triplet
HOMO $\rightarrow$ LUMO charge-transfer (CT) excitations
and to the corresponding singlet CT transitions.
A supermolecular basis was used in all cases. 
}
\label{fig2}
\end{figure}

In the following, we will investigate the applicability of (subsystem) TDDFT for 
the description of triplet excitation-energy transfer processes based on the helium 
dimer as an initial model system. 
Therefore, we present a comparison of the TEET and SEET descriptions 
using (i) approximate embedding (NAKE-sTDDFT), 
as well as (ii) ``exact'' embedding (PbE-sTDDFT).
Note that the NAKE-sTDDFT as well as the isolated calculations were performed using 
a supersystem basis. For the sake of completeness, it is important to mention that the 
excitation energies related to charge-transfer transitions of the analyzed excitation, 
stemming from the utilization of the supersystem basis, are also present in the results 
(as depicted in Fig.~\ref{fig2}). However, it is worth emphasizing that these 
excitation energies lack physical significance in the context of isolated and FDEu 
calculations~\cite{Jacob.2007} and therefore, will not be further discussed.

Fig.~\ref{fig2}a shows the excitation energies of the HOMO $\rightarrow$ LUMO triplet
(green lines) local excitation (LE) for an inter-subsystem separation of 1.0~\r{A} obtained with 
PbE-sTDDFT using the Fermi-shifted Huzinaga operator (see Appendix).
Moreover, the figure also illustrates the electronic couplings $V$ associated with 
this transition, which are directly related to the TEET coupling in a simple effective exciton model.
We have numerically verified that the triplet excitation energies of the 
helium dimer as obtained from supermolecular TDDFT calculations are reproduced 
using coupled PbE-sTDDFT. As a result, we omit the supersystem results for simplicity.

First, Fig.~\ref{fig2}a illustrates a comparison of TEET excitation energies 
obtained from different approaches: isolated TDDFT, 
FDEu-, and FDEc-TDDFT, revealing the impact of the embedding potential on 
the excitation energies. To analyze TEET effects, we first have to identify the 
corresponding triplet excitations. While this is straightforward for the isolated
system, already the FDEu results show triplet excitations of mixed local excitation
and charge-transfer character. For instance, the FDEu-PbE triplet excitation 
at 42.31~eV shows 62\% LE and 38\% CT character. The dominant orbital transitions of the 
triplet excitations are shown in \red{Tab.~S1} in the Supporting Information (SI), including 
the distinction between LE and CT transitions. Here, we decomposed the response of 
the system into local and CT contributions using the virtual-orbital-space 
localization method~\cite{Scholz.Tolle.2020}.
The FDEu-TDDFT vertical excitation energies show a noticeable decrease compared to the 
isolated case (43.33~eV), whereas the inclusion of inter-subsystem coupling effects 
leads to a splitting of FDEu energies into FDEc-TDDFT excitation energies at 38.17 and 
39.51~eV. These findings highlight the importance of the embedding approach in accurately 
describing excitation energies and demonstrates the importance of considering environmental 
effects and inter-subsystem interactions.

In the case of NAKE-sTDDFT, we investigate the excitation corresponding to the HOMO $\rightarrow$ 
LUMO LE transition considered in the PbE-sTDDFT calculations (see Fig.~\ref{fig2}b). 
The excitation energy decreases from 43.33~eV (isolated atom) to 42.88~eV when considering the 
active subsystem in the presence of the environmental subsystem. The splitting of the FDEu 
excitation energy into FDEc energies is significantly larger in PbE-sTDDFT compared to NAKE-sTDDFT.
Apart from the obvious difference in the treatment of the non-additive kinetic energy 
and external-orthogonality, also the mixing with charge-transfer excitations in the PbE case
may lead to differences in the resulting energy splittings (see \red{Tab.~S1} in the SI). 
Furthermore, the triplet excitation remains similarly stabilized in both NAKE- and PbE-sTDDFT 
methods, exhibiting minimal variation in energy between the two approaches.
With respect to the electronic coupling, the TEET coupling obtained from NAKE-sTDDFT (0.07~eV) 
differs from that of the PbE-sTDDFT calculation (1.43~eV) by about a factor of 20. 
This indicates that NAKE-sTDDFT significantly underestimates TEET electronic couplings, rendering 
it inadequate for accurately describing TEET couplings dominated by short-range effects. 

Fig.~\ref{fig2}c shows the corresponding singlet (red lines) HOMO $\rightarrow$ LUMO
LE transition of the helium dimer test system. Additionally to the triplet excitation, 
the dominant orbital transitions of the singlet excitations are shown in \red{Tab.~S2} in the SI.
For SEET, energies of 54.15~eV in the isolated case and 53.86~eV in the FDEu case are obtained 
with PbE which splits to 55.90 and 57.68~eV in the coupled FDE step. In the PbE case, the 
considered singlet transition shows approximately 80\% LE and 20\% CT character and a SEET 
coupling of 1.50~eV. 

When comparing the isolated and FDEu results, it is evident that the
reduction in energy for the singlet excitation is quite similar when using the PbE
and NAKE (53.94~eV) versions (see Fig.~\ref{fig2}d). 
In both cases, the singlet excitation energy is decreased by approximately 0.2 to 0.3~eV.  
Moreover, we observe that the SEET electronic coupling of the singlet HOMO $\rightarrow$ LUMO LE 
transition shows slightly lower values for NAKE- (0.43~eV) than for PbE-sTDDFT (1.50~eV).  
This indicates that both embedding methods provide a description of 
approximately equal quality of the SEET coupling for this specific transition, 
consistent with findings from other studies on SEET couplings~\cite{Toelle.2022}.
It is important to mention that we do not achieve identical results due to the
approximations employed in the NAKE-sTDDFT method.

When comparing the electronic coupling of triplet and singlet excitations, 
the situation is different for SEET couplings: Even though NAKE-sTDDFT yields a value
considerably lower than PbE-sTDDFT, the difference is not an order of
magnitude and can at least partially be explained by the use of different inter-subsystem
XC functionals, an aspect that will be further analyzed below.

\subsubsection{Fluoroethylene Dimer: Dependence on the Intra- 
and Inter-XC Approximations}\label{FDE_results}
As another example we analyze the HOMO($\pi$) $\rightarrow$ LUMO($\pi^*$) triplet 
excitation of a fluoroethylene dimer as a function of the inter-subsystem separation.
In this context, we investigate the influence of typical for the intra- and 
inter-subsystem XC functional as well as for the non-additive kinetic-energy functional
in NAKE-based subsystem TDDFT. 

\begin{figure}[!t]
\captionsetup[subfigure]{justification=centering}
 \centering
 \includegraphics[width=\textwidth]{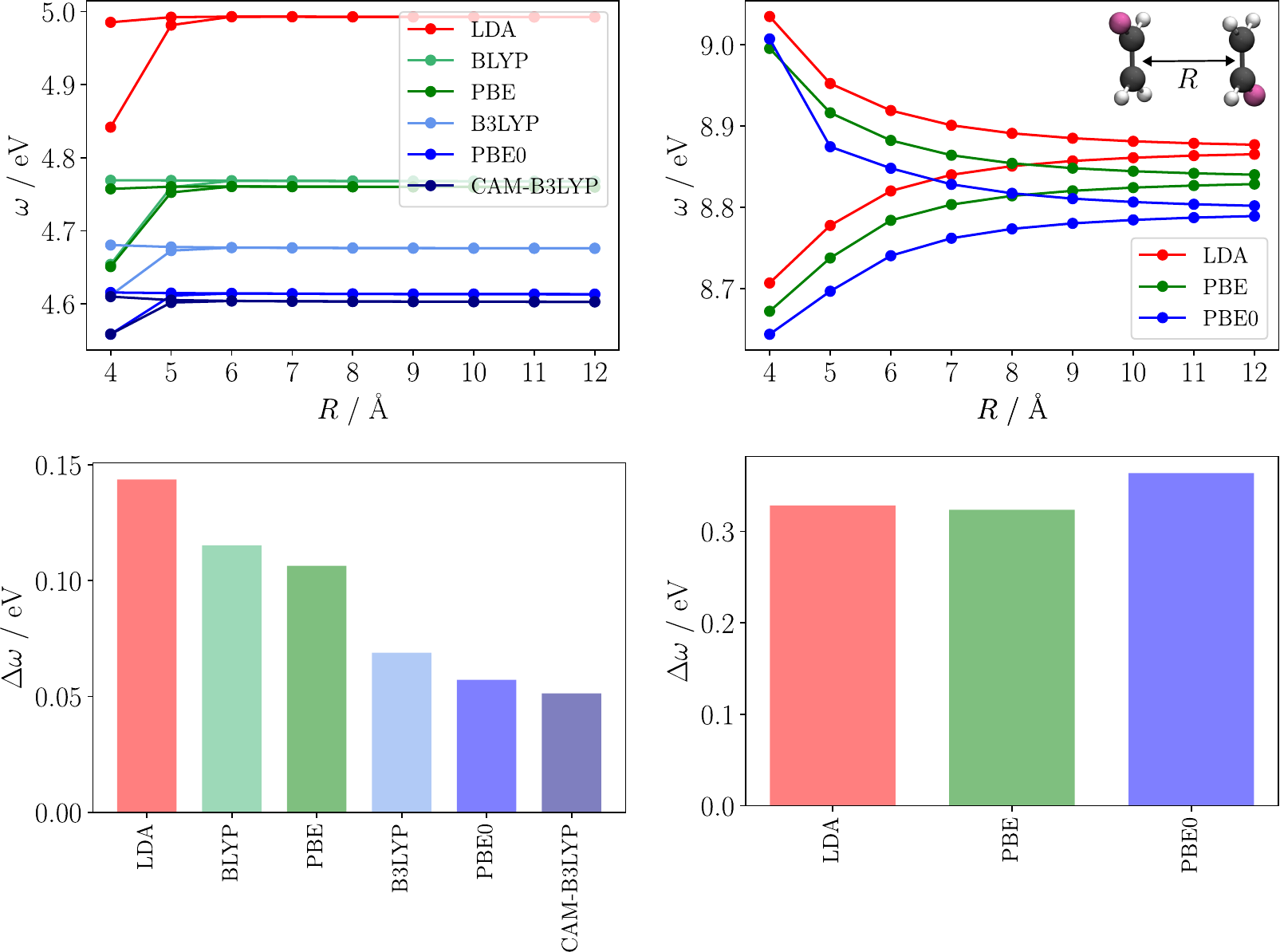}
    \begin{subfigure}[t]{0.49\textwidth}
        \caption{TEET}
        \label{fig:TEET}
    \end{subfigure}
    \begin{subfigure}[t]{0.49\textwidth}
        \caption{SEET}
        \label{fig:SEET}
    \end{subfigure}
\caption{
Excitation energies $\omega$ (top) 
of the $\pi \rightarrow \pi^*$ \subref{fig:TEET} triplet 
and \subref{fig:SEET} singlet excitation 
as a function of the inter-subsystem separation
and splittings $\Delta\omega$ (bottom) 
for an inter-subsystem separation of $R=4.0$~\r{A}
of the fluoroethylene dimer 
obtained from supersystem calculations
regarding the exchange--correlation functional dependence
[Supersystem: (LDA, BLYP, PBE, B3LYP, PBE0, CAM-B3LYP), def2-SVP].
}
\label{fig3}
\end{figure}

As a reference, we first study the corresponding excitation energies and splittings from
supermolecular TDDFT. In Fig.~\ref{fig3}, we depict the results obtained in these 
calculations employing various XC functionals. Excitation energies and splittings are shown
both for (a) triplet and (b) singlet excitations for inter-subsystem distance in the range 
from 4.0 to 12.0~\r{A}. As expected for a phenomenon governed by non-electrostatic, 
short-range effects, the splitting of the triplet excitation energies is significant only at
relatively short inter-subsystem distances, up until 6.0~\r{A}. Within this range, the 
splittings decrease rapidly with an increasing inter-subsystem separation.
The calculated splittings show a clear dependence on the XC functional:
The largest splitting (0.14~eV) is obtained with LDA, while GGAs result in splittings
between 0.12~eV (BLYP) and 0.11~eV (PBE). Hybrid functionals yield even lower splittings
of 0.07~eV (B3LYP), 0.06~eV (PBE0), and 0.05~eV (CAM-B3LYP).
Hence, the energy splitting significantly decreases with the amount of exact exchange 
included in the XC functional used. This behavior was also confirmed for other
systems (see \red{Sec.~S2} in SI).

For comparison, we also show the corresponding supermolecular results obtained for SEET
in Fig.~\ref{fig3}b, for which the dependence on the XC approximation is much 
smaller. Given that SEET is frequently governed by Coulomb interactions, the decrease in 
splitting occurs less rapidly with increasing distance than for TEET.
Furthermore, an asymmetric splitting behavior can be observed for TEET, 
in contrast to the SEET case. The contribution of the considered SEET transitions are 
shown in \red{Tab.~S3} in the SI.

\begin{figure}[!t]
\captionsetup[subfigure]{justification=centering}
\includegraphics[width=\textwidth]{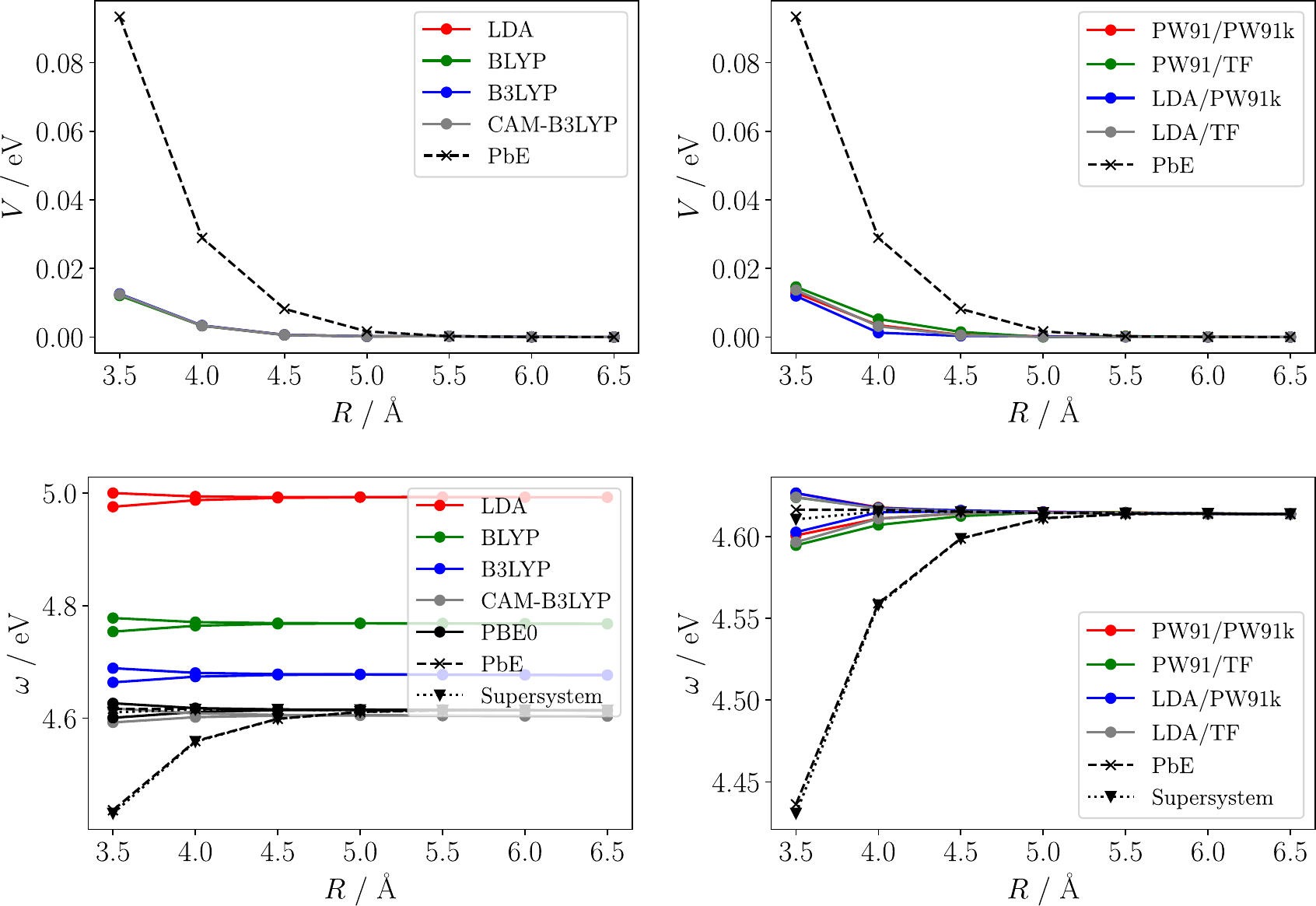}
 \centering
    \begin{subfigure}[c]{0.49\textwidth}
        \caption{}
        \label{fig:func}
    \end{subfigure}
    \begin{subfigure}[c]{0.49\textwidth}
        \caption{}
        \label{fig:basis}
    \end{subfigure}
\caption{
Electronic couplings $V$ (top) and excitation energies $\omega$ (bottom)
of the $\pi\rightarrow \pi^{*}$ triplet excitation
of the fluoroethylene dimer as a function of 
the inter-subsystem separation varying \subref{fig:func} the intra-subsystem
[NAKE: (LDA, BLYP, B3LYP, CAM-B3LYP, PBE0)/PW91/PW91k, def2-SVP]
and \subref{fig:basis} inter-subsystem XC functional dependence
[NAKE: PBE0/(PW91, LDA)/(PW91k, TF), def2-SVP]
with PbE and supersystem calculations as a reference
[PbE: PBE0/PBE0/level.; Supersystem: PBE0, def2-SVP].
}
\label{fig4}
\end{figure}

Turning now to the NAKE-sTDDFT calculations, we first investigated the effect of the 
intra-subsystem XC functional. Here, LDA~\cite{LDA}, BLYP~\cite{BLYP_1,BLYP_2}, PBE, 
B3LYP~\cite{B3LYP_1,B3LYP_2}, PBE0, and CAM-B3LYP~\cite{CAM-B3LYP} were considered for 
inter-subsystem distances between $R = 3.5$ and 6.5~\r{A}. The results are shown in
Fig.~\ref{fig4}a. 

While the excitation energies at larger distances vary roughly in the range between 4.6 
and 5.0~eV for the different intra-subsystem XC functionals, the splittings are very
similar and amount to approximately 0.025~eV for the shortest distance investigated here
(3.5~\r{A}). In a second step, we tested different combinations of inter-subsystem XC
(LDA, PW91) and NAKE (TF, PW91k) functionals. Changing these approximations does lead 
to changes in the splittings (see Fig.~\ref{fig4}b), but even the largest 
coupling obtained here (0.006~eV at 4.0~\r{A} for PW91/TF) is much smaller than the 
values obtained in PbE- or supermolecular calculations (see Fig.~\ref{fig3}a). 
Changing from a monomer to a supersystem basis does not lead to significantly different 
couplings in NAKE-sTDDFT calculations (see \red{Fig.~S1} in SI). This shows that with 
typical approximations as employed here, NAKE-sTDDFT is not well suited for the description 
of TEET because of the drastic underestimation of the TEET electronic couplings.

\subsection{PbE-based Subsystem TDDFT for TEET Couplings}\label{PbE_results}
While full equivalence to supermolecular results can be achieved with PbE-sTDDFT, this 
embedding approach also allows for the introduction of approximations regarding the choice 
of approximate intra- and inter-subsystem XC functionals, as well as variations in the projection operator 
(see \red{Sec.~S4} in SI). A particular focus will be on the influence of exact exchange. 
Moreover, an analysis of the different TEET kernel contributions will be carried out
to assess their relative importance.

\subsubsection{The Role of Exact Exchange}\label{XX}
\begin{figure}[t]
\centering
\includegraphics[width=\textwidth]{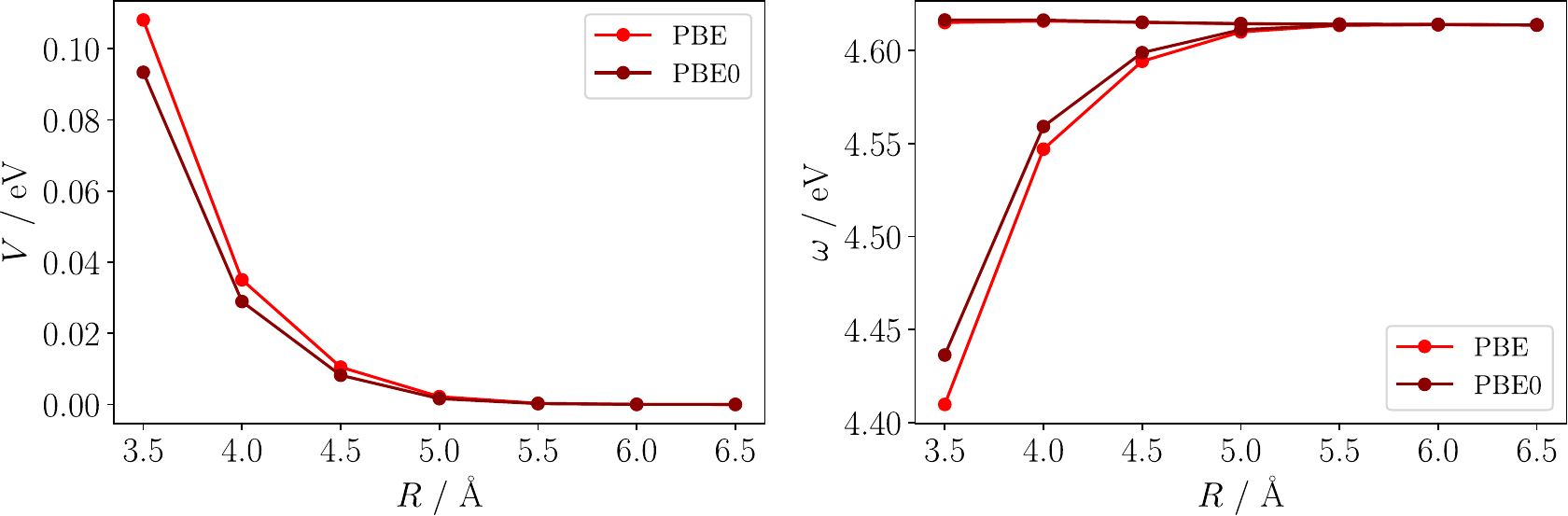}
\caption{
Electronic couplings $V$ (left) and excitation energies $\omega$ (right) 
of the $\pi \rightarrow \pi^*$ triplet excitation of the fluoroethylene dimer 
as a function of the inter-subsystem separation 
with and without exact inter-subsystem exchange 
[PbE: PBE0/(PBE0, PBE)/level., def2-SVP].
}
\label{fig5}
\end{figure}

In order to further investigate the origin of the XC functional dependence observed in 
supermolecular TDDFT, the calculations from Sec.~\ref{FDE_results} were repeated using 
PbE-sTDDFT while varying the XC as well as the non-additive XC functional 
to differentiate between intra- and inter-subsystem XC approximations. 
Hence, we examined all combinations of the PBE0 and PBE functionals for the intra- and 
inter-subsystem interaction to assess how exact exchange affects the electronic couplings 
and excitation energies of the fluoroethylene dimer. As expected, when consistently varying
both XC contributions at the same time, i.e. PBE/PBE or PBE0/PBE0 (see \red{Fig.~S2a} in 
the SI), we observe the same changes as in the supermolecular case 
(cf. Fig.~\ref{fig3}). However, changing the amount of exact inter-subsystem exchange
does not significantly affect the excitation energies and splittings: As shown in 
Fig.~\ref{fig5}, the results obtained with PBE0/PBE are very similar to the PBE0/PBE0 
data. Similarly, the PBE/PBE0 data are very close to the PBE/PBE results (see \red{Fig.~S2a} 
in the SI).

These results also demonstrate that exact exchange is not necessary in the inter-subsystem 
XC approximation to obtain hybrid-functional-quality TEET couplings. 
This has important consequences on the overall scaling behavior of this approach, 
as the expensive formal $n^4$-scaling when using hybrid functionals can be restricted to 
the intra-subsystem part.

\subsubsection{Separating the Different Kernel Contributions} 
\label{TEETKernel}
\begin{figure}[!t]
\centering
\includegraphics[width=\textwidth]{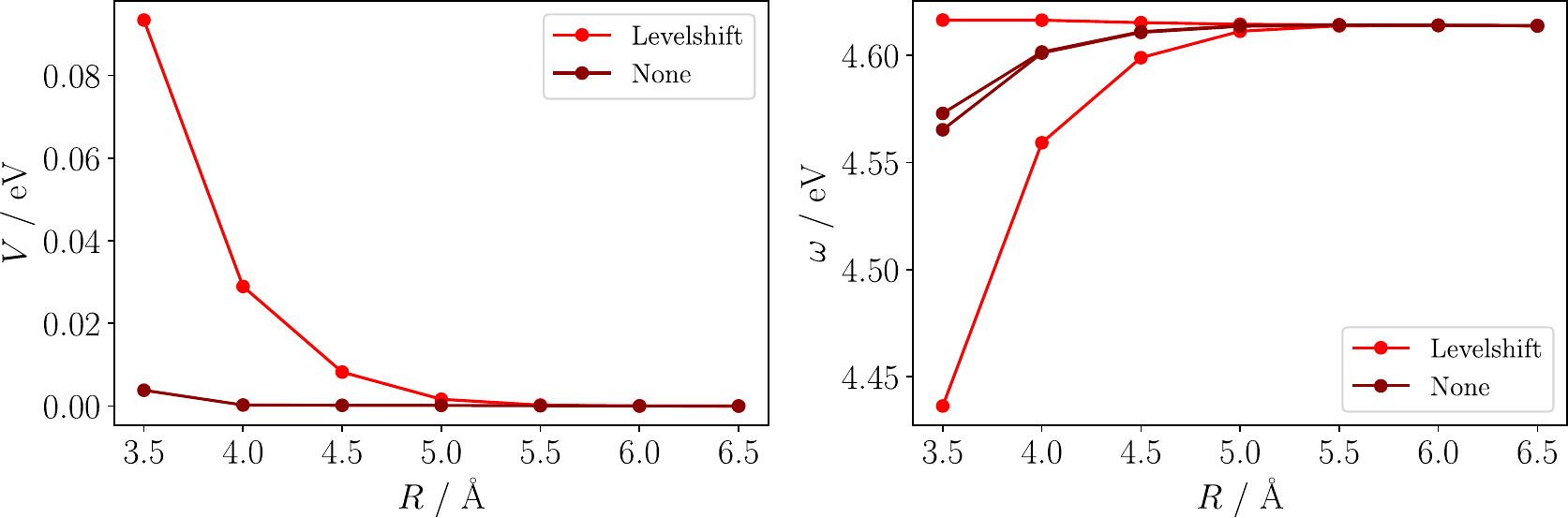}
\caption{
Electronic couplings $V$ (left) and excitation energies $\omega$ (right) 
of the $\pi\rightarrow \pi^{*}$ triplet excitation as a function
of the inter-subsystem separation of the fluoroethylene dimer 
with and without considering the EO kernel
[PbE: PBE0/PBE0/(level., None), def2-SVP].
}
\label{fig6}
\end{figure}

To gain a deeper understanding of the different contributions to the TEET couplings, 
we analyzed the individual contributions to the corresponding coupling matrix elements. 
Assuming the same approximation is used for all intra-subsystem XC functionals and the
non-additive XC functional, the PbE-sTDDFT triplet coupling matrix using the levelshift
operator is defined as 
\begin{equation}
    K^\mathrm{(-), level.}_{(ia)_I, (jb)_J} 
    = \left(\phi_i^I\phi_a^I\left| f\msb{xc}^{\alpha\alpha} 
    - f\msb{xc}^{\alpha\beta} \right|\phi_j^J\phi_b^J\right) 
    + c\msb{HF}(\phi_i^I\phi_j^J|\phi_a^I\phi_b^J)
    + \mu \langle\phi_{i}^I|\phi_{j}^J\rangle \langle\phi_{b}^J|\phi_{a}^I\rangle,
\label{eq:TEETCoupling}
\end{equation}
where the first part of this equation describes the non-additive XC (NAXC) kernel, 
the second the exact exchange (XX) kernel, and 
the third the external-orthogonality (EO) contribution. 
Fig.~\ref{fig6} presents results obtained with and without the EO kernel contribution. 
These results reveal that, when considering the EO kernel, 
the TEET coupling at an inter-subsystem separation of 4.0~\r{A} is approximately 0.03~eV, 
accompanied by an energy splitting of 0.06 eV.
Upon removal of the EO kernel, the electronic coupling virtually vanishes, 
corresponding to energy splittings close to zero.

\begin{figure}[!t]
\centering
    \includegraphics[width=0.5\textwidth]{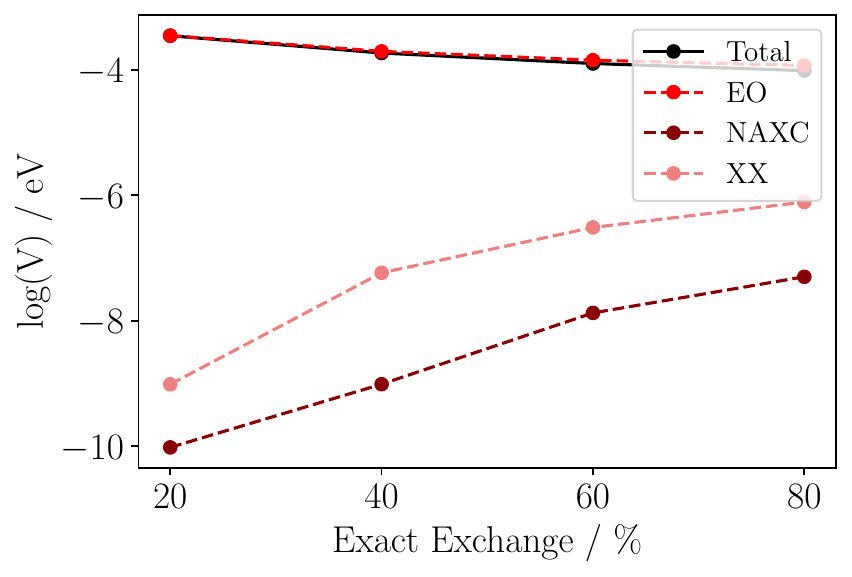}  
\caption{
Total electronic coupling $V$,  as well as contributions of the external--orthogonality (EO), 
non-additive exchange--correlation (NAXC), and exact exchange (XX) kernel,
of the $\pi \rightarrow \pi^*$ triplet excitation
for an inter-subsystem separation of 4.0~\r{A} 
of the fluoroethylene dimer
obtained from PbE-sTDDFT calculations 
as a function of the amount of exact exchange $c\msb{HF}$
in the intra- and inter-subsystem exchange--correlation functional. 
PBEx refers to PBE functionals 
where \textit{x} denotes the amount of exact exchange in percent
[PbE: PBEx/PBEx/level., def2-SVP].
}
\label{fig7}
\end{figure}

The results presented in Fig.~\ref{fig7} provide further evidence for the importance 
of the EO kernel contributions. In that figure, we show the three different 
contributions to the TEET couplings identified in Eq.~\eqref{eq:TEETCoupling} as a function 
of the amount of exact exchange (both in the intra- and inter-subsystem contributions).
It can be seen that the TEET coupling is dominated by the EO kernel independently of the
amount of exact exchange, while the NAXC and XX contributions are at least two to three 
orders of magnitude smaller. Considering that the EO kernel depends significantly on the 
inter-subsystem orbital overlap, the dependence of TEET couplings on the XC functional can 
most likely be attributed to the orbital topology rather than the inter-subsystem XC 
approximation. To ensure that the choice of the levelshift parameter $\mu$ does not 
influence TEET couplings when using the levelshift operator version of PbE, we systematically
varied this parameter in the range of $10^2$ to $10^7$. No significant changes in the electronic
couplings could be observed within this parameter range (see Tab.~S7 in the SI).
Consequently, we can confidently assert that the 
dependence on the intra-subsystem XC functional is likely related to the topology
of the molecular orbitals involved in the transition. 
A slight decrease of the contribution of the EO kernel, and consequently the TEET coupling 
kernel can be observed in Fig.~\ref{fig7} as the amount of exact intra-subsystem
exchange increases. This can also be attributed to the dependence of the EO kernel on orbital 
overlap, as a lower amount of exact exchange in the intra-subsystem XC functional leads to 
less compact and more delocalized molecular orbitals (see, e.g., Refs.~\citenum{Yang.2008,
Burke.2014,Burke.2019,Massolle.2020}).
As a result, the occupied orbitals need to be more strongly localized in the PbE 
approach, requiring larger orbital rotations to enforce orthogonality of the 
inter-subsystem orbitals. Therefore, a smaller amount of exact intra-subsystem exchange 
results in larger TEET couplings.

\subsection{Application: Investigation of Solvent Effects on TEET Couplings 
of PDI Dimer Systems} \label{PDI_Dimer}
While NAKE-sTDDFT may not be well suited for a proper description of direct TEET 
couplings as outlined above, it may still be useful to describe explicit solvent
effects on TEET in combination with PbE-sTDDFT.
Hence, in the following, we will analyze solvent effects on TEET couplings.
The mixed PbE-/NAKE-embedding approach introduced in Sec.~\ref{HybridEmbedding}, 
which should be able to take environmental effects into account
without compromising accuracy of the direct coupling between the active (PbE-)systems. 

As a test system, we choose a (solvated) PDI dimer system. 
PDI plays an important role in diverse studies concerning potential 
applications in the field of optoelectronic devices~\cite{PDI.OSC.2013}, such as organic 
solar cells and light-emitting diodes and nanotechnology~\cite{PDI.Nano.2022} as well as 
drug delivery~\cite{PDI.DrugDelivery.2021} due to its self-assembly behavior. 
Recently, PDI derivatives have been discussed as possible substituents for fullerenes as 
the electron acceptor in organic solar cells~\cite{PDI.OSC.2017}.
Moreover, PDI molecules have been found to exhibit high electron mobility 
and efficient charge transport properties, 
rendering PDI-based materials promising for organic electronic devices.
In addition to its electronic properties, PDI monomers can form a variety of 
self-assembled structures, including micelles, fibers, and gels.
For our purposes, the most important properties of PDI dimer systems are their 
robustness, modifiable absorption spectrum range, appreciable extinction coefficients, 
and tendency for strong $\pi-\pi$ interactions~\cite{PDI.2022}.
As such they constitute suitable candidates for benchmarking the different 
embedding methods for TEET couplings.

\subsubsection{Environmental Effect of Explicit Solvent Molecules} 
\label{SolventEffect}
\begin{figure}[!t]
    \centering
    \includegraphics[width=0.7\textwidth]{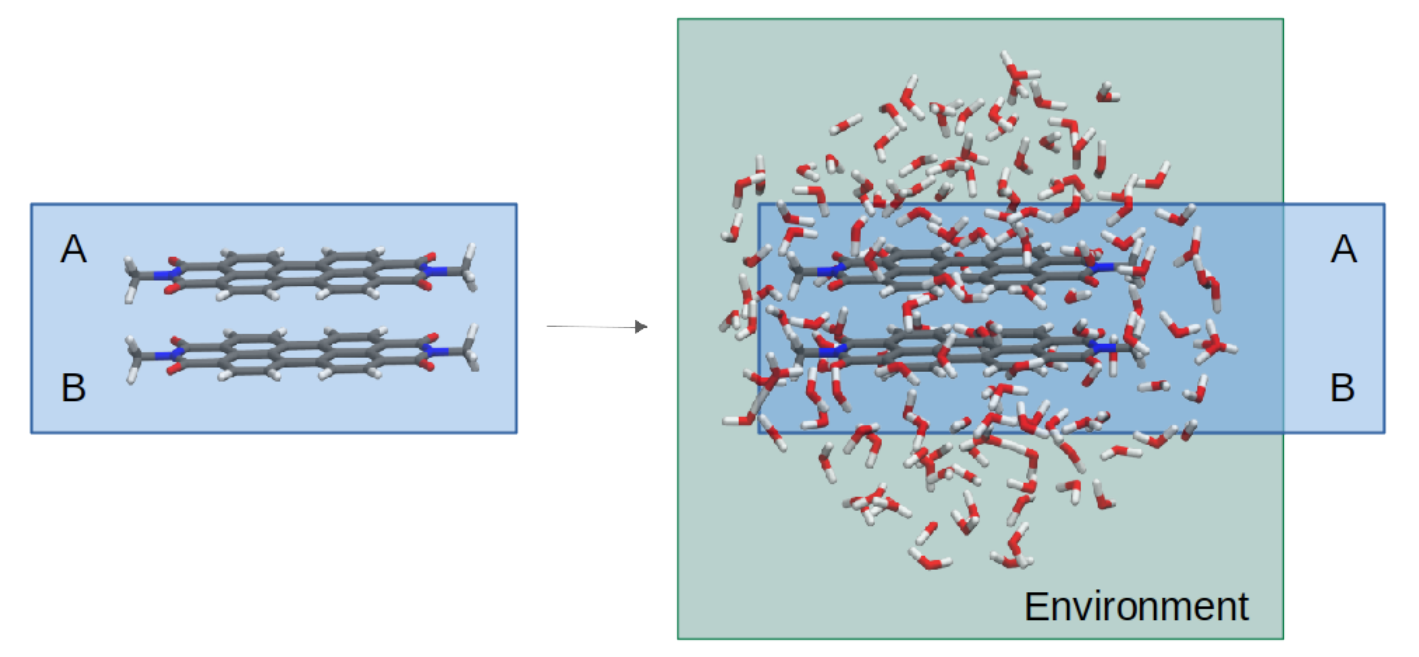}
    \caption{Illustration of the mixed PbE-/NAKE-embedding scheme
    considering the structure of the PDI dimer surrounded by 149 water molecules,
    taken from Ref.~\citenum{PDI.Struc.Curutchet.2009}. Description of the 
    PDI dimer using PbE highlighted in blue and the environmental 
    NAKE region highlighted in green.}
    \label{fig8}
\end{figure}

\begin{figure}[!t]
\captionsetup[subfigure]{justification=centering}
\centering
\includegraphics[width=\textwidth]{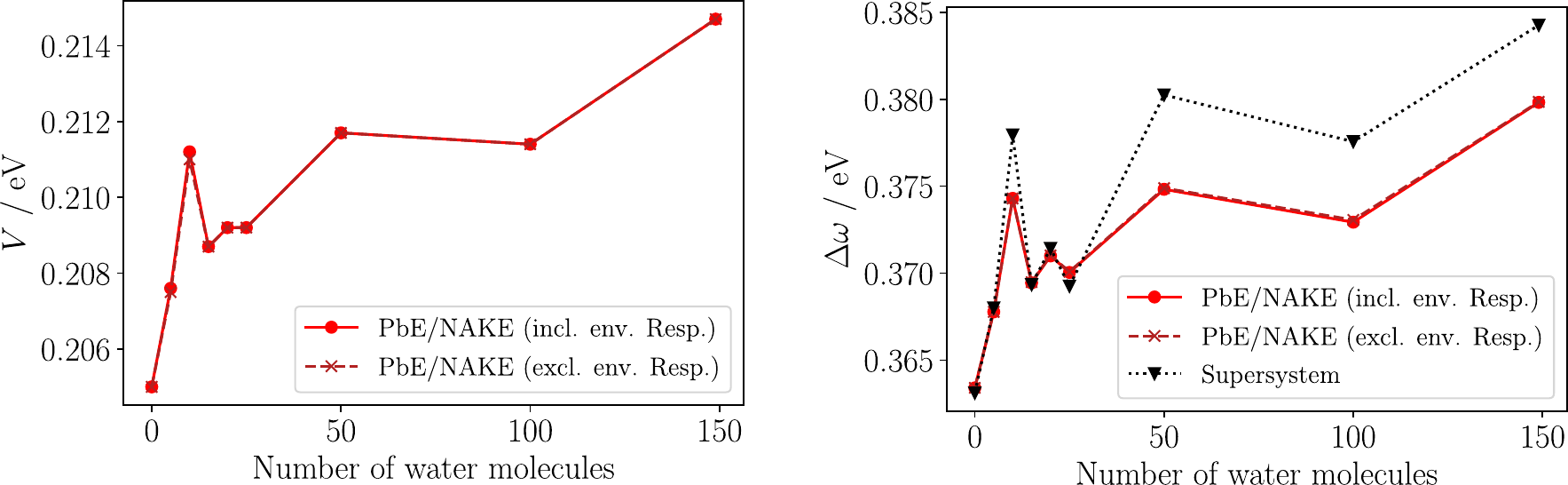}
    \begin{subfigure}[c]{0.49\textwidth}
        \caption{}
        \label{fig:Hybrid}
    \end{subfigure}
    \begin{subfigure}[c]{0.49\textwidth}
        \caption{}
        \label{fig:Hybrid_Super}
    \end{subfigure}
\caption{
\subref{fig:Hybrid} Electronic couplings $V$ 
and \subref{fig:Hybrid_Super} splittings $\Delta\omega$
of the $\pi \rightarrow \pi^*$ triplet excitations
for an inter-subsystem separation of 3.5~\r{A}
of the PDI dimer in water  
varying the number of water molecules 
[Supersystem: PBE0;
PbE/NAKE: [PBE0, PBE]/[PBE0, PBE]/[level., PW91K], def2-SVP].
}
\label{fig9}
\end{figure}

As a first test, we investigate the TEET coupling in a solvated PDI dimer with fixed 
distance as a function of the size of the solvation shell
(see Fig.~\ref{fig8}).
The six lowest-energy $\pi \rightarrow \pi^*$ triplet excitations of the PDI dimer 
were determined for an inter-subsystem separation of 
3.5~\r{A}, while varying the number of water molecules (see Fig.~\ref{fig9}).
The selection of water molecules for this purpose was based on the distance from the center
of mass of each water molecule to the nearest atom of either of the PDI molecules.
The calculations discussed in this subsection were performed with version 1.6.0 of the \textsc{Serenity} program\cite{serenity.zenodo.1.6.0}.

The results presented in Fig.~\ref{fig9} show (a) electronic couplings between the PDI molecules and 
(b) energy splittings obtained with the mixed PbE-/NAKE-embedding approach, in comparison
to the corresponding supersystem splittings as a function of the number of water molecules 
included. 
In the mixed PbE-/NAKE-embedding calculations, the electronic structures were relaxed in three initial freeze-and-thaw cycles (considering the dimer as a single subsystem). Potentials and kernels in FDE and FDEu calculations for either of the monomers employed a numerical integration grid extending over both molecules (due to the PbE-based description of their interaction). FDE and FDEu calculations for the water molecules used monomer grids and the final FDEc calculation employed a supersystem grid to determine inter-subsystem couplings.
For the mixed PbE-/NAKE-embedding approach, we show two sets of data. 
For the first set, we included \emph{all} response couplings between the PDI molecules (whose interaction
is described with PbE) as well as the environmental water molecules (whose interaction is described with NAKE functionals). Within the approximations introduced in the mixed embedding approach,
this is equivalent to a supersystem treatment as it includes excitations of and couplings between all subsystems.
For the second set, while still including the coupling between the PDI molecules, we neglected the couplings to
excitations on the environmental water molecules expecting them to be small.
The vacuum supersystem splitting is 0.363~eV which increases to 0.384~eV in 
the presence of the complete solvation shell containing 149 water molecules.
Especially for smaller numbers of water molecules, this behavior shows some irregularities
with intermediate decreases in the splitting as the solvation shell grows. Concerning the 
PbE/NAKE results, the mixed embedding method qualitatively reproduces the supersystem 
results, although, for cases involving a larger number of water molecules, there is a 
deviation of approximately 5~meV. These discrepancies can most likely be traced back to 
the fact that the response of the solvent molecules is not considered explicitly in this
approximate embedding approach. 
We find no appreciable difference in the electronic couplings computed with the mixed 
embedding approach, providing evidence that triplet couplings based on NAKE functionals are
negligibly small, in line with the findings of previous sections.
Therefore, for all subsequent mixed PbE-/NAKE-embedding calculations, the
explicit response of the solvent molecules is omitted, leading to a computational advantage.
Regarding the TEET couplings acquired through the PbE/NAKE approach, a similar trend
emerges in relation to changes in the number of solvent molecules, as observed for the 
energy splittings. Note that in this system, the TEET electronic couplings are not simply
given as half of the energy excitation splitting $\Delta\omega$. The reason is that this
simple relation does not hold for systems with non-symmetric monomers as is the case here 
because of the inclusion of explicit solvent molecules.
Overall, the environmental effect of the solvent molecules surrounding the chromophores 
of interest is non-negligible but rather small compared to the total splitting. 
This is in line with the observations made 
by Curutchet and Voityuk~\cite{PDI.Curutchet.JPC.2012}.

\subsubsection{Bridge Effect of Explicit Solvent Molecules} \label{BridgeEffect}
In Ref.~\citenum{PDI.Curutchet.JPC.2012}, it turned out that solvent molecules located between
the interacting chromophores can more strongly affect the TEET couplings. In the following,
we will investigate this aspect in the context of subsystem TDDFT. To this end, we analyze 
the distance-dependence of TEET couplings in a solvated, stacked PDI dimer in water and two 
organic solvents of different polarity, namely chloroform (Chl) and benzene (Bnz).
In contrast to the example studies above, these structures feature solvent molecules
between the PDI monomers, which may enhance the TEET in terms of bridge-mediated
effects~\cite{PDI.Curutchet.JPC.2012}.
The structures of these systems were taken from Ref.~\citenum{PDI.Curutchet.JPC.2012} 
with inter-subsystem distances of $R = 7.0$ and 10.5~\r{A}, taking ten snapshots for each 
solvent and each distance into consideration. Moreover, additional calculations for a 
PDI--PDI structure with a separation of $R=3.5$~\r{A} were performed to generate reference 
values for evaluating TEET couplings without bridging solvent molecules. 

\begin{figure}[!t]
\captionsetup[subfigure]{justification=centering}
\centering
    \includegraphics[width=\textwidth]{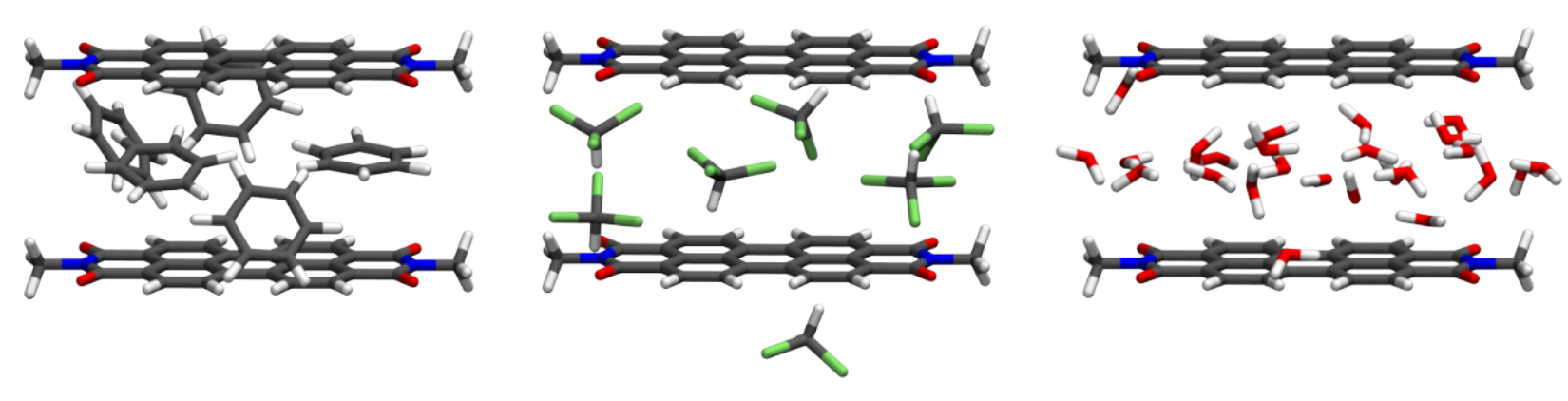}
    \begin{subfigure}[b]{0.33\textwidth}
        \caption{Benzene}
    \end{subfigure}
    \begin{subfigure}[b]{0.33\textwidth}
        \caption{Chloroform}
    \end{subfigure}
    \begin{subfigure}[b]{0.32\textwidth}
        \caption{Water}
    \end{subfigure}
\caption{Structures of the PDI dimer with 25 water, 
seven chloroform, and five benzene molecules in between.}
\label{fig10}
\end{figure}

\begin{figure}[t]
    \centering
    \includegraphics[width=0.5\textwidth]{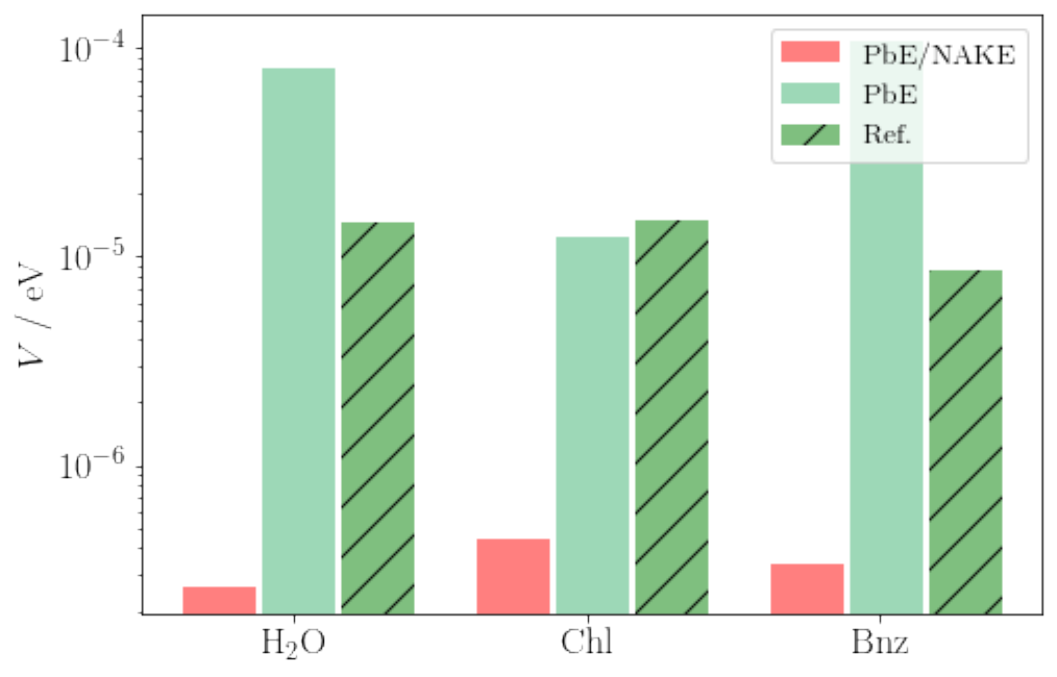}
\caption{
TEET couplings $V$ at $R = 7.0$~\r{A} inter-subsystem separations 
of the $\pi \rightarrow \pi^*$ triplet excitations
of the PDI dimer in different solvents 
calculated with mixed PbE-/NAKE-embedding and PbE-sTDDFT
with respect to the bridge effect of the solvent molecules 
[PbE: PBE0/PBE/level.;
PbE/NAKE: [PBE0, PBE]/[PBE, PBE]/[level., PW91K], def2-SVP].
}
\label{fig11}
\end{figure}

Initial calculations were performed using mixed PbE-/NAKE-embedding for a fixed 
inter-subsystem separation of $R = 7.0$~\r{A}, corresponding to 25 water, seven chloroform, 
and five benzene molecules, respectively. Only one snapshot for each solvent was computed
(see structures in Fig.~\ref{fig10}). As can be seen from 
Fig.~\ref{fig11}, the resulting TEET couplings differ by two orders of 
magnitude from the reference data of Ref.~\citenum{PDI.Curutchet.JPC.2012}.
This shows that the mixed PbE-/NAKE-embedding is not suitable for describing the 
bridge-mediated coupling effect of the solvent molecules located in between 
the chromophores.

For this reason, the response effect of the bridging solvent molecules has to be 
included explicitly in the description of the electronic couplings using PbE-sTDDFT. 
This can be achieved by defining two subsystems, each consisting of one chromophore and 
a certain number of solvent molecules. The latter was determined by calculating the distance 
from the center of mass of each solvent molecule to the nearest atom of each PDI monomer
and subsequently, assigning each of the solvent molecules to one of the monomers 
based on the solvent--monomer distances. We have also confirmed that different assignments 
of the solvent molecules to the subsystems only have a minimal effect on the obtained TEET couplings.
Including the bridge effect of the solvent molecules in PbE-sTDDFT calculations, 
TEET couplings at a similar order of magnitude as in Ref.~\citenum{PDI.Curutchet.JPC.2012} 
were obtained (see Fig.~\ref{fig11}). For instance, when considering 25 water 
molecules, we determined a TEET coupling of $8.01 \times 10^{-5}$~eV for the 
HOMO $\rightarrow$ LUMO transition of the PDI dimer at $R=7.0$\r{A}, while Curutchet and 
Voityuk reported a coupling of $1.47 \times 10^{-5}$~eV for the same 
snapshot~\cite{PDI.Curutchet.JPC.2012}. It has to be pointed out that Curutchet and 
Voityuk used the Fragment Excitation Difference (FED) method~\cite{Hsu.2008}
for computing the couplings, so that some remaining differences ca be 
expected~\cite{Tolle.2020}.

\subsubsection{Conformational Sampling of TEET Couplings}

\begin{figure}[t]
    \centering
    \includegraphics[width=0.7\textwidth]{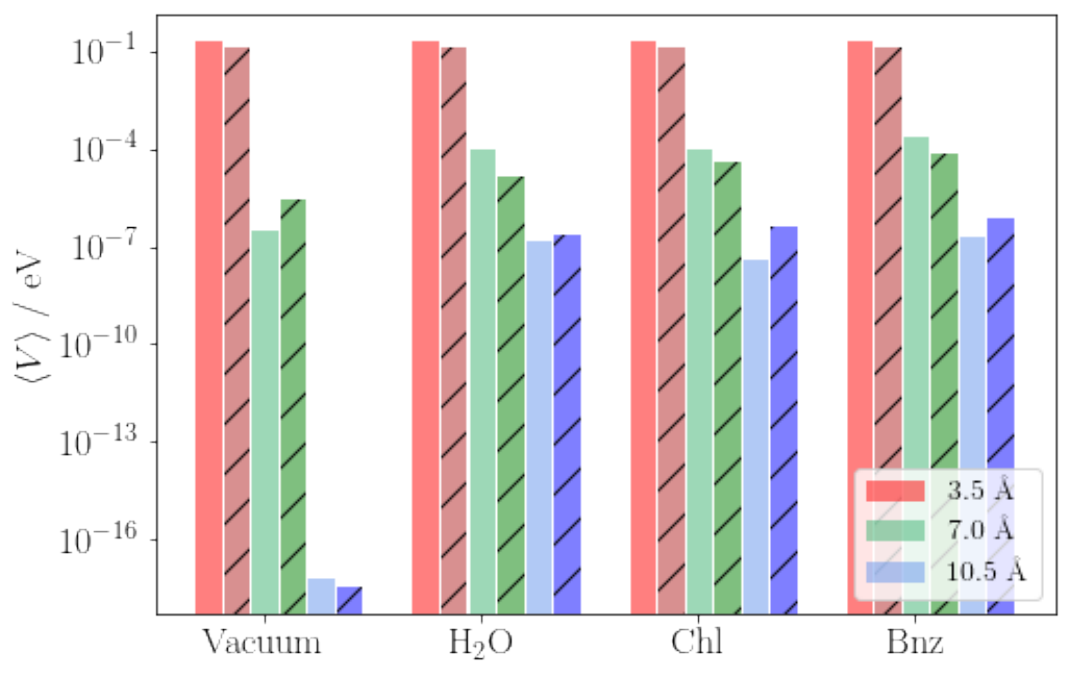}
\caption{
Average TEET couplings $\langle V \rangle$
at $R = 3.5$, $7.0$, and $10.5$~\r{A} inter-subsystem separations 
of the $\pi \rightarrow \pi^*$ triplet excitations
of the PDI dimer in benzene, chloroform, and water solutions 
calculated with PbE-sTDDFT (solid)
[PbE: PBE0/PBE/level., def2-SVP],
compared to the results in Ref.~\citenum{PDI.Curutchet.JPC.2012} (striped).
}
\label{fig12}
\end{figure}

The solvent-mediated coupling strongly depends on the relative orientation of the 
bridging solvent molecules because of the dependence of TEET couplings on the orbital 
overlap. Therefore, small changes in the orientation of the solvent or the chromophores
can have a large impact on the TEET coupling values. Hence, we calculated the couplings 
for ten selected chromophore--solvent structures~\cite{PDI.Curutchet.JPC.2012} differing 
in the relative orientation of the solvent molecules in order to provide a more realistic 
description of the solvent effect. From the data shown in Fig.~\ref{fig12}, 
it can be observed that the average TEET coupling $\langle V \rangle$ at an inter-subsystem 
separation of $R=10.5$~\r{A} is around $10^{-18}$~eV in a vacuum environment. However, when 
considering the system in solution, the TEET coupling increases to approximately $10^{-7}$ 
to $10^{-8}$~eV for all three solvents considered here, which is qualitatively in agreement 
with Ref.~\citenum{PDI.Curutchet.JPC.2012}. At an 
inter-subsystem separation of 7.0~\r{A}, the influence of the solvent leads to
an increase of the TEET couplings to about $10^{-4}$ to $10^{-5}$~eV, compared 
to the TEET couplings of roughly $10^{-7}$~eV in vacuum. In absolute terms, this is a much
stronger increase in coupling, while the relative increase is only $10^3$ compared 
to an increase of $10^{11}$ for an inter-subsystem separation of 10.5~\r{A}.
These results unambiguously demonstrate that the bridge effect of solvent molecules is 
crucial when characterizing medium- or long-range TEET processes. 

\begin{figure}[t]
\centering
        \includegraphics[width=0.5\textwidth]{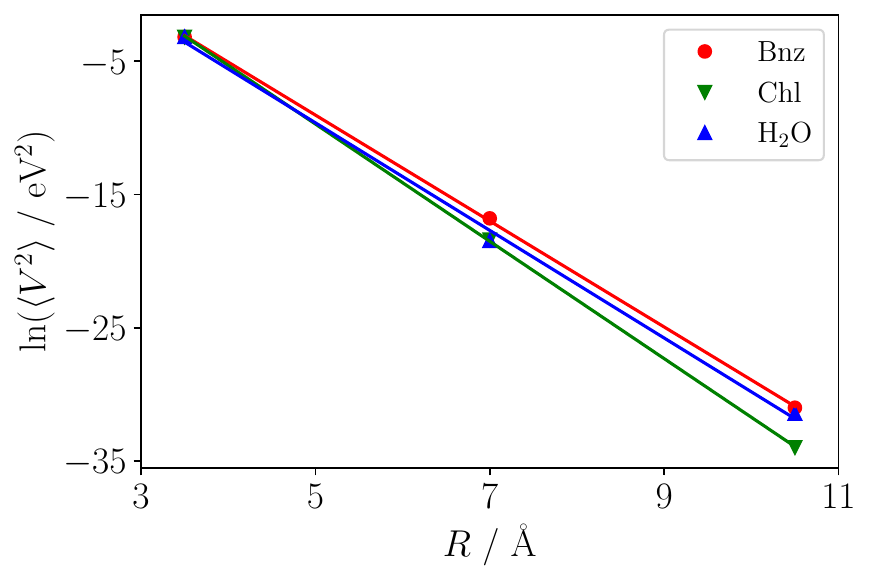}
\caption{
Logarithm of average squared electronic couplings $\langle V^2 \rangle$
of the $\pi \rightarrow \pi^*$ triplet excitations 
of the PDI dimer at $R = 3.5$, 7.0, and 10.5~\r{A} inter-subsystem separations 
in benzene, chloroform, and water solutions calculated with PbE-sTDDFT
[PbE: PBE0/PBE/level., def2-SVP].
} 
\label{fig13}
\end{figure}

Comparing the calculated couplings for different solvents as shown in 
Fig.~\ref{fig13}, a decrease of the solvent-mediated coupling from benzene via 
chloroform to water can be observed. Curutchet and Voityuk by contrast observed a reduction 
in the TEET coupling when going from benzene via chloroform to water. In order to further elucidate 
the differences between our results and those of Ref.~\citenum{PDI.Curutchet.JPC.2012},
we conducted a recalculation of TEET couplings for ten structures 
of the PDI dimer system with 25 water molecules located in between and an inter-subsystem
distance of $R=7.0$~\r{A} (as illustrated in Fig.~\ref{fig10}).
We employed the FED approach starting from TDA [PBE0/def2-SVP] and CIS [HF/def2-SVP]
results for determining the electronic couplings and compared the results
to both the original data from Ref.~\citenum{PDI.Curutchet.JPC.2012} [HF/6-31+G(d)] 
and to our results obtained with PbE-sTDDFT [PBE0/PBE/level., def2-SVP]. 
The TEET couplings (see \red{Tab.~S8} in the SI) from FED based on TDA are in good agreement
with PbE-sTDDFT, while FED results based on CIS calculations are in good agreement with 
the data from Ref.~\citenum{PDI.Curutchet.JPC.2012}. This shows that the discrepancies between
our study and Ref.~\citenum{PDI.Curutchet.JPC.2012} primarily stem from the choice of the 
electronic structure method (HF vs KS-DFT with PBE0), rather than from the approach used to
calculate the TEET couplings (PbE-sTDDFT vs. FED).

\section{Conclusion}
While subsystem Time-Dependent Density-Functional Theory (sTDDFT) based Non-Additive 
Kinetic Energy (NAKE) functionals is suitable for describing Singlet Excitation-Energy 
Transfer (SEET), we have demonstrated in this study that it is not well-suited for 
investigating Triplet Excitation-Energy Transfer (TEET). 
The energy splittings associated with TEET are severely underestimated when using 
sTDDFT with approximate NAKE functionals, deviating significantly from supersystem 
calculations. Projection-based Embedding (PbE) eliminates these problems and offers a 
way to directly extract electronic couplings as it acts as a diabatization scheme. 
The combination of projection-based embedding with sTDDFT showed promising results for 
accurately describing TEET processes as long as bridge-mediated TEET is not significant. 

A more detailed analysis of the applicability of PbE-sTDDFT for TEET couplings 
showed that TEET couplings increase as the amount of exact exchange in the 
XC functional decreases. The functional dependence was found to originate from the 
intra-subsystem XC functional, while the inter-subsystem XC functional does not have
a significant impact on the electronic couplings. 
We have demonstrated that inter-subsystem exact exchange is not necessary to 
obtain reliable TEET couplings, and can be replaced by (semi-)local DFT-exchange.
This has important consequences when studying TEET couplings in aggregates of 
coupled chromophores, potentially offering a large performance advantage, 
as the expensive calculation of exact exchange can be restricted to within the subsystems. 
A comprehensive study of the composition of the TEET coupling kernel showed that 
the external-orthogonality (EO) kernel plays a dominant role.
As the amount of exact intra-subsystem exchange increases, the significance of the
EO kernel, and consequently the TEET coupling kernel, decreases. 
This can be attributed to the EO kernel's dependence on orbital overlap, 
as a higher amount of exact exchange in the intra-subsystem XC functional 
results in more compact and localized molecular orbitals.
Hence, it can be assumed that the functional dependence of TEET couplings 
on the intra-subsystem XC approximation originates from the molecular orbital structure.

The investigation of solvated PDI dimer systems revealed that the mixed 
PbE-/NAKE-embedding approach introduced in Sec.~\ref{HybridEmbedding} offers the ability 
to consider environmental effects on TEET between pigments in close contact. However, 
this approach is no longer applicable if bridge-mediated effects become important. 
Bridge-mediated TEET can be accurately described by explicitly including the bridging 
fragments in the PbE-sTDDFT treatment. 
Regarding the conformational sampling, our results have shown that while the obtained TEET 
couplings for different solvents and inter-subsystem separations are comparable to the data in 
Ref.~\citenum{PDI.Curutchet.JPC.2012}, the observed trend with respect to the solvents is different. 
Curutchet and Voityuk reported a decrease in electronic couplings from benzene via chloroform 
to water, while our calculations indicate a decrease from benzene to water to chloroform. 
We identified the electronic structure method as the origin of these discrepancies
(CIS vs. PBE0/PBE), rather than the approach for calculating the TEET couplings:
Both PbE-sTDDFT and FED are suitable for calculating TEET couplings in bridge-mediated 
donor-acceptor systems and yield similar results if the same electronic-structure method 
is chosen. Subsystem TDDFT can offer an efficiency advantage in this context for more 
complex and larger systems, as it avoids the need for extensive supersystem calculations. 
This can pave the way for investigating TEET phenomena in realistic systems as involved in
photocatalysis, photovoltaics, and optoelectronic devices, including the effect of a medium 
surrounding the chromophore groups.

\begin{acknowledgement}
Funding by the Deutsche Forschungsgemeinschaft (DFG, German Research Foundation) 
through SFB 1459 (Project A03 -- 433682494) is gratefully acknowledged.
We would like to thank Carles Curutchet for providing the perylene diimide dimer geometries
and the data used in \red{Tab.~S8} in the SI.
J.T. gratefully acknowledges funding by the Deutsche Forschungsgemeinschaft 
(DFG, German Research Foundation) through DFG-495279997.
\end{acknowledgement}

\bibliography{literature}

\section{Appendix}
\subsection{EO Kernel of the Fermi-Shifted Huzinaga Potential}
We apply the following definitions for notational brevity in the following:
\begin{align}
  F^{IJ}_{\mu \nu} &= \langle \chi_\mu^I |\hat F| \chi_\nu^J\rangle, \\
  S^{IJ}_{\mu \nu} &= \langle \chi_\mu^I        | \chi_\nu^J\rangle,
\end{align}
where $I$ and $J$ label subsystems, $\hat F$ is the supersystem Fock operator, and $\chi$ are basis functions, which can, but need not necessarily, belong to the same supermolecular basis set.
The Fermi-shifted Huzinaga potential\cite{Chulhai.2018} in the atomic-orbital basis of subsystem $A$ in the presence of another subsystem $B$ reads:
\begin{align}
    V_{\mu\nu}^{\mathrm{Fermi}, A(B)} &= -\sum_{\kappa\lambda}\left(F^{AB}_{\mu\kappa} - \epsilon_F S^{AB}_{\mu\kappa}\right) D^B_{\kappa\lambda} S^{BA}_{\lambda\nu}
                  -\sum_{\kappa\lambda} S^{AB}_{\mu\kappa} D^{B}_{\kappa\lambda} \left(F^{BA}_{\lambda\nu} - \epsilon_F S^{BA}_{\lambda\nu}\right) \\
               &= -\sum_{\kappa\lambda} F^{AB}_{\mu\kappa} D^B_{\kappa\lambda} S^{BA} _{\mu\nu} - \sum_{\kappa\lambda}S^{AB}_{\mu\kappa} D^{B}_{\kappa\lambda} F^{BA}_{\mu\nu} + 2\epsilon_F\sum_{\kappa\lambda} S^{AB}_{\mu\kappa} D^B_{\kappa\lambda} S^{BA}_{\mu\nu}\\
               &= V_{\mu\nu}^{\mathrm{Huz.}, A(B)} + 2\epsilon_F\sum_{\kappa\lambda} S^{AB}_{\mu\kappa} D^B_{\kappa\lambda} S^{BA}_{\mu\nu},
\end{align}
where the first two summands are equivalent to the regular Huzinaga potential\cite{Hegely.2016} and the second one resembles the potential arising from the use of the levelshift operator\cite{Manby.Miller.2012} scaled with the Fermi shift $\epsilon_F$.
To obtain an expression of the corresponding coupling-matrix contributions of 
$V_{\mu\nu}^{\mathrm{Fermi}, A(B)}$, we first rewrite the density matrix of $B$ in the molecular-orbital representation,
\begin{equation}
    D_{\kappa\lambda}^B = \sum_{(pq)_B} C^B_{\kappa p} D^B_{pq}  C^{B}_{\lambda q}
\end{equation}
and transform the whole expression to the molecular-orbital basis of $A$
\begin{align}
    V_{ia}^{\mathrm{Fermi}, A(B)} &= \sum_{\mu\nu} C_{\mu i}^A V_{\mu\nu}^{\mathrm{Fermi}, A(B)} C_{\nu a}^A \\
    &= - \sum_{(pq)_B} F^{AB}_{ip} D^{B}_{pq} S^{BA}_{qa} 
                                        - \sum_{(pq)_B} S^{AB}_{ip} D^{B}_{pq} F^{BA}_{qa}
                                        + 2\epsilon_F \sum_{(pq)_B} S^{AB}_{ip} D^{B}_{pq} S^{BA}_{qa} \\
    &= V_{ia}^{\mathrm{Huz.}, A(B)} + 2\epsilon_F \sum_{(pq)_B} S^{AB}_{ip} D^{B}_{pq} S^{BA}_{qa}.
\end{align}
The external-orthogonality kernel of the Fermi-Shifted Huzinaga operator is then obtained as
\begin{align}
    K_{(ia)_A,(jb)_B}^{\mathrm{Fermi}, A(B)} &= \frac{\partial V_{ia}^{\mathrm{Fermi}, A(B)}}{\partial D^{B}_{jb}} \\
    &= K_{(ia)_A,(jb)_B}^{\mathrm{Huz.}, A(B)} + 2\epsilon_F S^{AB}_{ij} S^{BA}_{ba},
\end{align}
where the contribution beyond the regular Huzinaga kernel\cite{Toelle.2019b} (which, in turn, resembles the levelshift kernel contribution\cite{Neugebauer.2019}) may become non-negligible if no supersystem basis set is used (which would imply that $S^{AB}_{ij} = S^{BA}_{ba} = 0$). We, therefore, explicitly include it in calculations with the Fermi-shifted Huzinaga operator.

\end{document}


\vspace*{0cm}
\begin{center}
{\LARGE
Triplet Excitation-Energy Transfer Couplings 
from Subsystem Time-Dependent Density-Functional Theory
}
\vspace{2cm}

{\Large -- Supporting Information --}
\vspace{2cm}

{\large
Sabine K\"{a}fer$^{1}$,
Niklas Niemeyer$^{1}$, 
Johannes T\"{o}lle$^{2}$,
and Johannes Neugebauer$^{1,}$\footnote{email: j.neugebauer@uni-muenster.de}}
\\[1cm]

$^1$University of M\"unster, 
Organisch-Chemisches Institut \\
and Center for Multiscale Theory and Computation,\\
Corrensstra{\ss}e 36, 48149 M\"unster, Germany

$^2$Division of Chemistry and Chemical Engineering, \\
California Institute of Technology, Pasadena, California 91125, USA
\end{center}
\vfill

\begin{tabular}{ll}
Date: & December 12, 2023 \\
\end{tabular}
\thispagestyle{empty}

\clearpage

\section{Dominant Orbital Transitions}\label{OrbTransitions}

\subsubsection*{a) Helium Dimer}
\begin{table}[H]
\caption{
Excitation energies $\omega$ and dominant orbital transitions of the four lowest-lying 
triplet excitations of the helium dimer at 1.0~\r{A} separation, 
corresponding to the HOMO $\rightarrow$ LUMO local excitation (LE) 
and charge-transfer (CT) transitions. 
}
\centering
\begin{tabular}{c c c c c c}
\hline
\hline
  State & $\omega$ / eV & System & Type & Dominant orbital transition \\
\hline
\hline
  T$_1$ & 37.71 & 1 & LE & HOMO $\rightarrow$ LUMO (31.35\%) \\
        &       & 2 & LE & HOMO $\rightarrow$ LUMO (31.35\%) \\
        &       & 1 & CT & HOMO $\rightarrow$ LUMO (18.65\%) \\
        &       & 2 & CT & HOMO $\rightarrow$ LUMO (18.65\%) \\ \hline
  T$_2$ & 39.19 & 1 & LE & HOMO $\rightarrow$ LUMO (33.24\%) \\
        &       & 2 & LE & HOMO $\rightarrow$ LUMO (33.24\%) \\
        &       & 1 & CT & HOMO $\rightarrow$ LUMO (16.76\%) \\
        &       & 2 & CT & HOMO $\rightarrow$ LUMO (16.76\%) \\ \hline
  T$_3$ & 49.25 & 1 & CT & HOMO $\rightarrow$ LUMO (33.24\%) \\
        &       & 2 & CT & HOMO $\rightarrow$ LUMO (33.24\%) \\
        &       & 1 & LE & HOMO $\rightarrow$ LUMO (16.76\%) \\
        &       & 2 & LE & HOMO $\rightarrow$ LUMO (16.76\%) \\ \hline
  T$_4$ & 50.49 & 1 & CT & HOMO $\rightarrow$ LUMO (31.35\%) \\
        &       & 2 & CT & HOMO $\rightarrow$ LUMO (31.35\%) \\
        &       & 1 & LE & HOMO $\rightarrow$ LUMO (18.65\%) \\
        &       & 2 & LE & HOMO $\rightarrow$ LUMO (18.65\%) \\ 
\hline
\hline
\end{tabular}
\label{Tab:HeDimer_OrbTransitions_TEET}
\end{table}

\begin{table}[H]
\caption{
Excitation energies $\omega$ and dominant orbital transitions of the four lowest-lying 
singlet excitations of the helium dimer at 1.0~\r{A} separation, 
corresponding to the HOMO $\rightarrow$ LUMO local excitation (LE) 
and charge-transfer (CT) transitions. }
\centering
\begin{tabular}{c c c c c c}
\hline
\hline
  State & $\omega$ / eV & System & Type & Dominant orbital transition \\
\hline
\hline
  S$_1$ & 43.25 & 1 & CT & HOMO $\rightarrow$ LUMO (40.15\%) \\
  &       & 2 & CT & HOMO $\rightarrow$ LUMO (40.15\%) \\
  &       & 1 & LE & HOMO $\rightarrow$ LUMO ( 9.85\%) \\ 
  &       & 2 & LE & HOMO $\rightarrow$ LUMO ( 9.85\%) \\ \hline
        
  S$_2$ & 43.95 & 2 & CT & HOMO $\rightarrow$ LUMO (41.17\%) \\
  &       & 1 & CT & HOMO $\rightarrow$ LUMO (41.17\%) \\
  &       & 1 & LE & HOMO $\rightarrow$ LUMO ( 8.83\%) \\ 
  &       & 2 & LE & HOMO $\rightarrow$ LUMO ( 8.83\%) \\ \hline 
        
  S$_3$ & 55.44 & 1 & LE & HOMO $\rightarrow$ LUMO (41.17\%) \\
  &       & 2 & LE & HOMO $\rightarrow$ LUMO (41.17\%) \\
  &       & 1 & CT & HOMO $\rightarrow$ LUMO ( 8.83\%) \\
  &       & 2 & CT & HOMO $\rightarrow$ LUMO ( 8.83\%) \\ \hline
        
  S$_4$ & 57.26 & 2 & LE & HOMO $\rightarrow$ LUMO (40.15\%) \\
  &       & 1 & LE & HOMO $\rightarrow$ LUMO (40.15\%) \\
  &       & 1 & CT & HOMO $\rightarrow$ LUMO ( 9.85\%) \\
  &       & 2 & CT & HOMO $\rightarrow$ LUMO ( 9.85\%) \\
\hline
\hline
\end{tabular}
\label{Tab:HeDimer_OrbTransitions_SEET}
\end{table}

\subsubsection*{b) Fluoroethylene Dimer}
\begin{table}[H]
\caption{
Excitation energies $\omega$ and dominant orbital transitions of the four lowest-lying 
triplet excitations of the fluoroethylene dimer at 4.0~\r{A} separation, 
corresponding to the HOMO $\rightarrow$ LUMO local excitation (LE) 
and charge-transfer (CT) transitions. 
}
\centering
\begin{tabular}{c c c c c c}
\hline
\hline
  State & $\omega$ / eV & System & Type & Dominant orbital transition \\
\hline
\hline
  T$_1$ &  4.56 & 2 & LE & HOMO $\rightarrow$ LUMO (48.57\%) \\
        &       & 1 & LE & HOMO $\rightarrow$ LUMO (48.56\%) \\ \hline
  T$_2$ &  4.62 & 1 & LE & HOMO $\rightarrow$ LUMO (49.62\%) \\
        &       & 2 & LE & HOMO $\rightarrow$ LUMO (49.62\%) \\ \hline
  T$_3$ &  7.19 & 1 & CT & HOMO $\rightarrow$ LUMO (49.99\%) \\
        &       & 2 & CT & HOMO $\rightarrow$ LUMO (49.99\%) \\ \hline
  T$_4$ &  7.26 & 2 & CT & HOMO $\rightarrow$ LUMO (48.93\%) \\
        &       & 1 & CT & HOMO $\rightarrow$ LUMO (48.92\%) \\  
\hline
\hline
\end{tabular}
\label{Tab:FluoroethlyeneDimer_OrbTransitions}
\end{table}

\section{Dependence of TEET Couplings on Intra- and Inter-Subsystem XC Approximations}
\label{FuncDependence}

\subsubsection*{a) DNA Base Pairs}
\begin{table}[H]
\caption{
Excitation energies $\omega$ (in eV) and splittings $\Delta\omega$ (in eV)
of the $\pi \rightarrow \pi^*$ triplet excitation
of the stacked adenine--adenine DNA base pair in A- and B-DNA
obtained from supersystem calculations with different exchange--correlation functionals.
Two excitation energies are given because of the splitting of triplet excitations.
Side (left) and front (right) view of stacked A--A DNA bases in A- and B-DNA
[Supersystem: (LDA, BLYP, PBE, B3LYP, PBE0, CAM-B3LYP), def2-SVP].
}
\begin{figure}[H]
    \begin{minipage}[c]{0.24\textwidth}     
        \begin{center}
            \includegraphics[width=\textwidth]{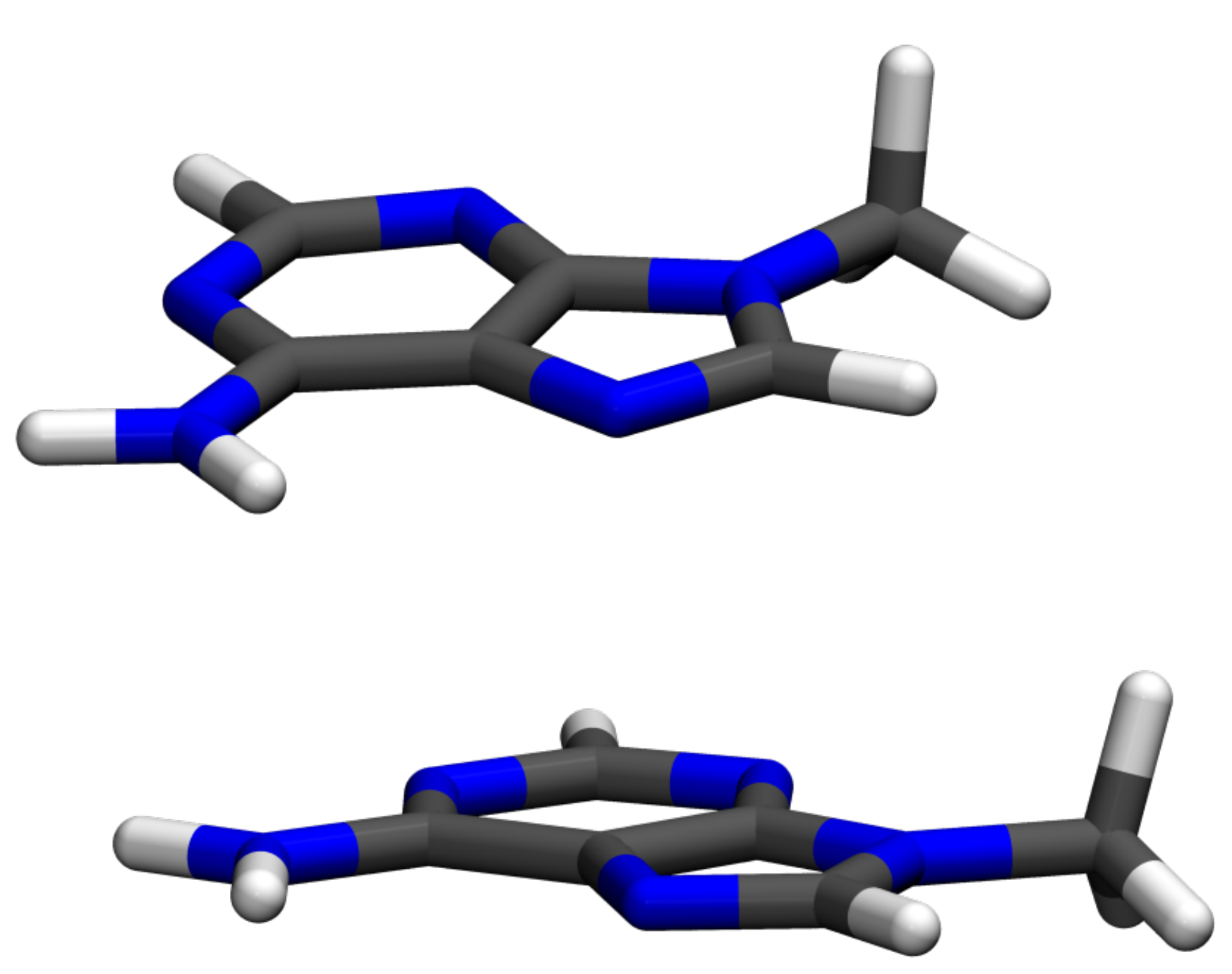}
        \end{center}
    \end{minipage}
    \begin{minipage}[c]{0.24\textwidth}     
        \begin{center}
            \includegraphics[width=\textwidth]{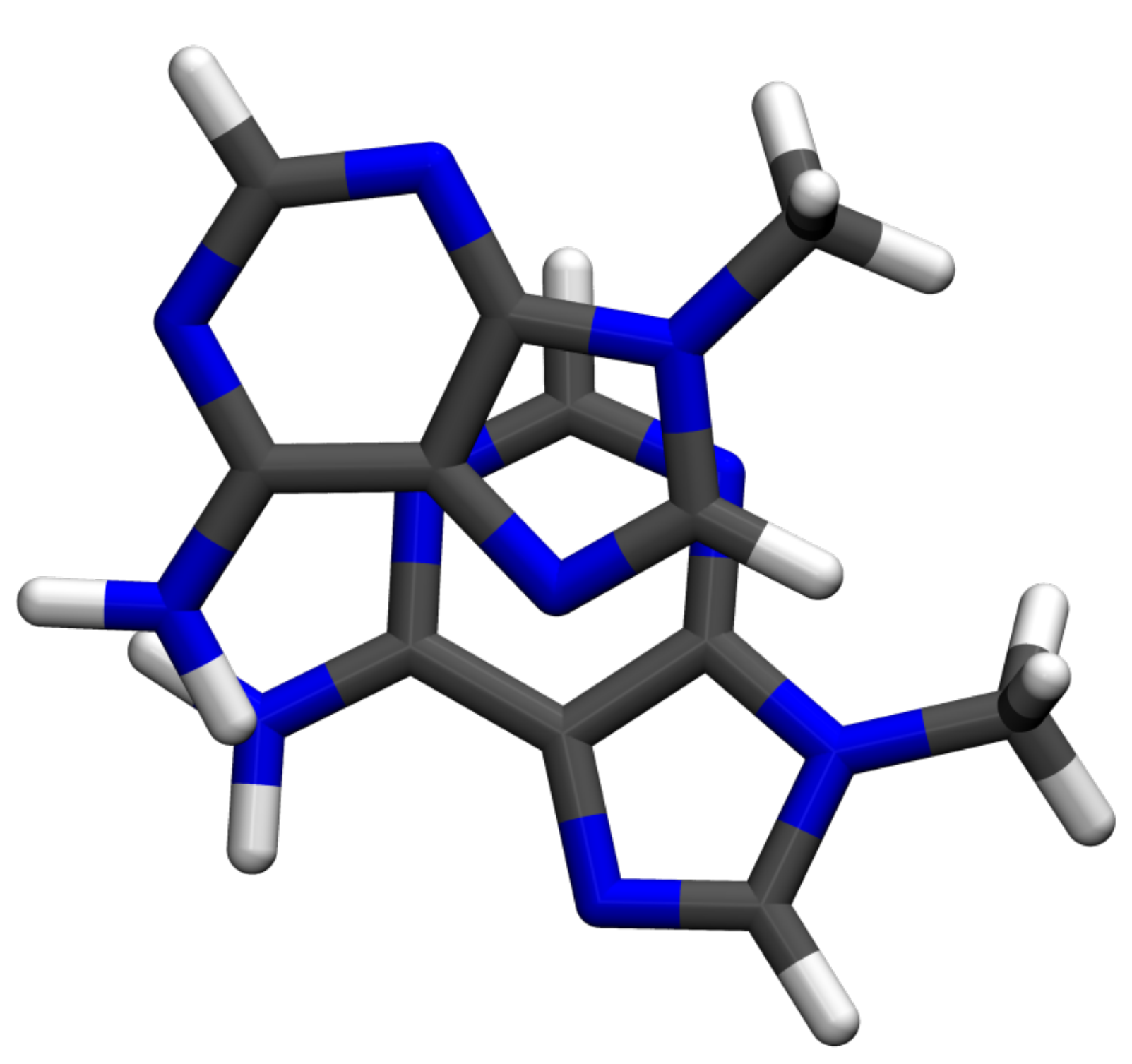}
        \end{center}
    \end{minipage}
\vline
    \begin{minipage}[c]{0.24\textwidth}     
        \begin{center}
            \includegraphics[width=0.9\textwidth]{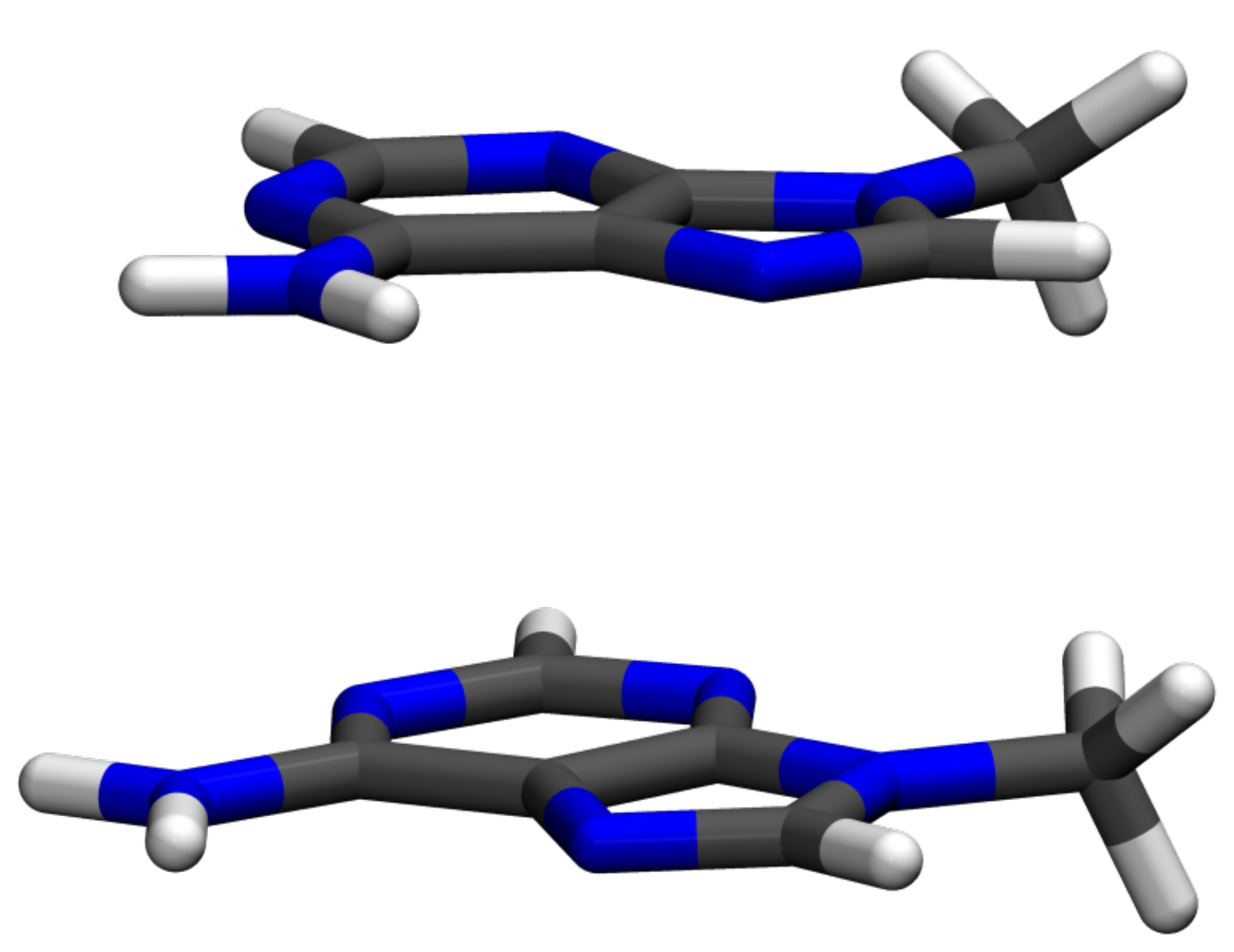}
        \end{center}
    \end{minipage}
    \begin{minipage}[c]{0.24\textwidth}     
        \begin{center}
            \includegraphics[width=0.9\textwidth]{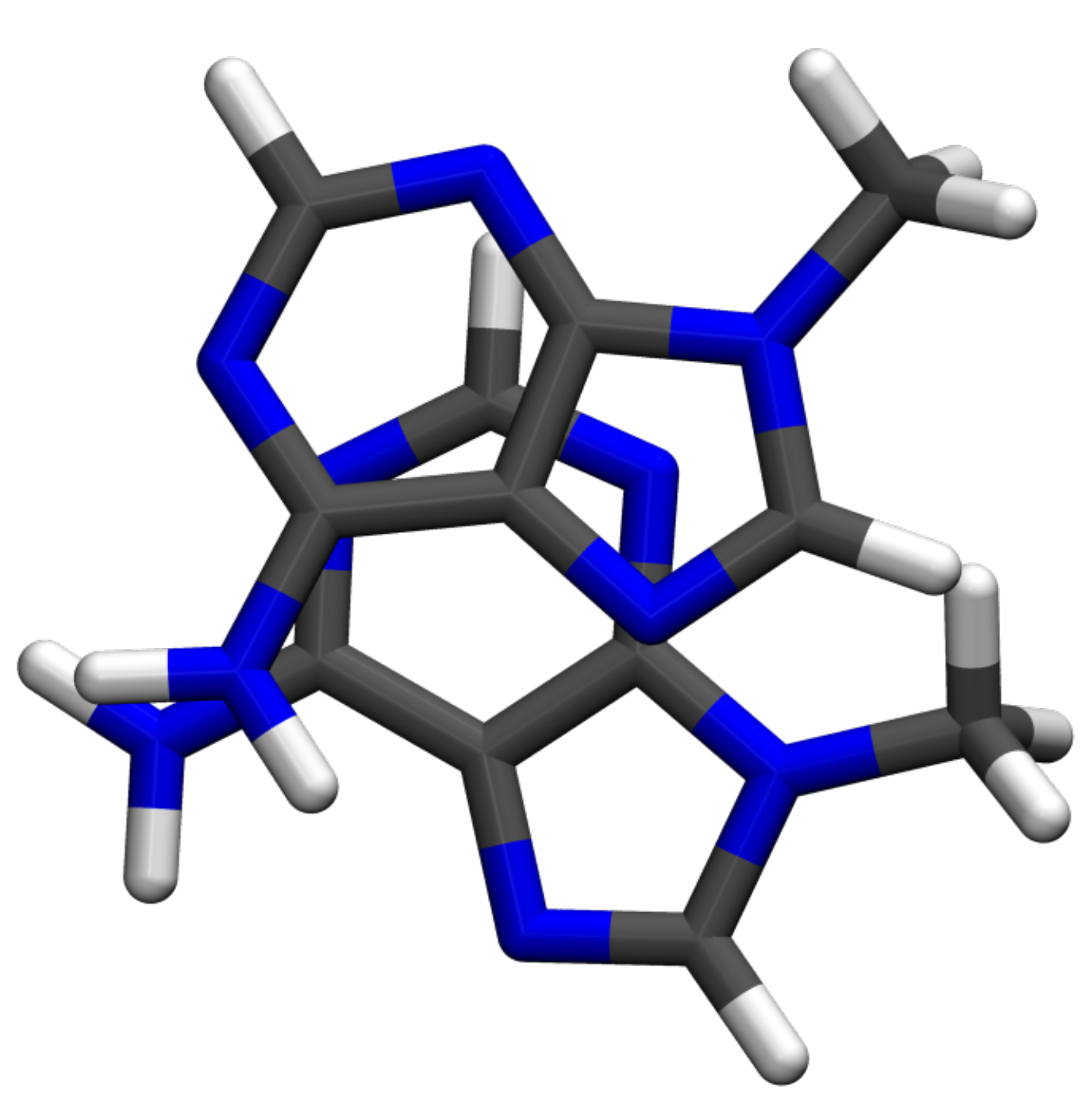}
        \end{center}
    \end{minipage}   
\vspace{0.2cm}   
    \begin{minipage}[c]{0.49\textwidth}     
        \begin{center}
            \footnotesize{A-DNA} 
        \end{center}
    \end{minipage}
    \begin{minipage}[c]{0.49\textwidth}     
        \begin{center}
            \footnotesize{B-DNA} 
        \end{center}
    \end{minipage}
\end{figure}
\centering
\begin{tabular}{l l l l l l l}
\hline
\hline
        & \multicolumn{6}{c}{XC Functional} \\
        & LDA & BLYP & PBE & B3LYP & PBE0 & CAM-B3LYP \\
\hline
        \textbf{A-DNA} & & & & & & \\
        $\omega$ & 3.62766 & 3.58628 & 3.59648 & 3.79983 & 3.83253 & 3.93740 \\
              & 3.72505 & 3.65340 & 3.66270 & 3.82437 & 3.85131 & 3.94868 \\
\cline{2-7}
$\Delta\omega$ & 0.09739 & 0.06712 & 0.06622 & 0.02454 & 0.01878 & 0.01128 \\ 
\hline
        \textbf{B-DNA} & & & & & & \\
        $\omega$ & 3.71373	& 3.64155 & 3.65195 & 3.81453 & 3.84312 & 3.94336 \\
              & 3.75848 & 3.67793 & 3.68842 & 3.84303 & 3.86979 & 3.96660 \\
\cline{2-7}
$\Delta\omega$ & 0.04475 & 0.03638 & 0.03647 & 0.02850 & 0.02667 & 0.02324 \\ 
\hline
\hline
\end{tabular}
\label{tab:AA_DNABasePair_xcFunc}
\end{table}

\begin{table}[H]
\caption{
Excitation energies $\omega$ (in eV) and splittings $\Delta\omega$ (in eV)
of the $\pi \rightarrow \pi^*$ triplet excitation 
of the stacked thymine--thymine DNA base pair in A- and B-DNA
obtained from supersystem calculations 
with different exchange--correlation functionals.
two excitation energies are given because of the splitting of triplet excitations.
Side (left) and front (right) view of stacked T--T DNA bases in A- and B-DNA
[Supersystem: (LDA, BLYP, PBE, B3LYP, PBE0, CAM-B3LYP), def2-SVP].
}
\begin{figure}[H]
    \begin{minipage}[c]{0.24\textwidth}     
        \begin{center}
            \includegraphics[width=\textwidth]{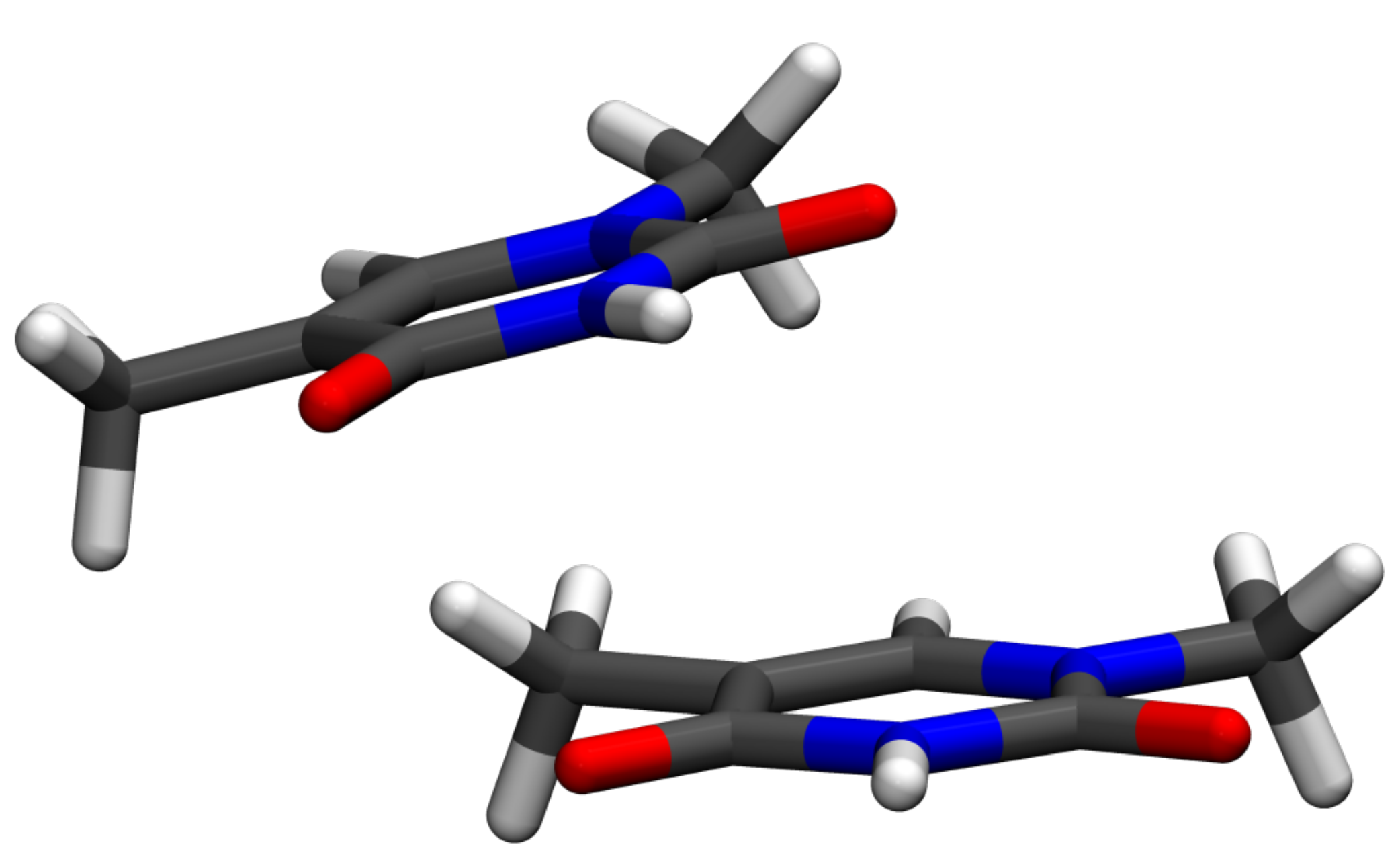}
        \end{center}
    \end{minipage}
    \begin{minipage}[c]{0.24\textwidth}     
        \begin{center}
            \includegraphics[width=\textwidth]{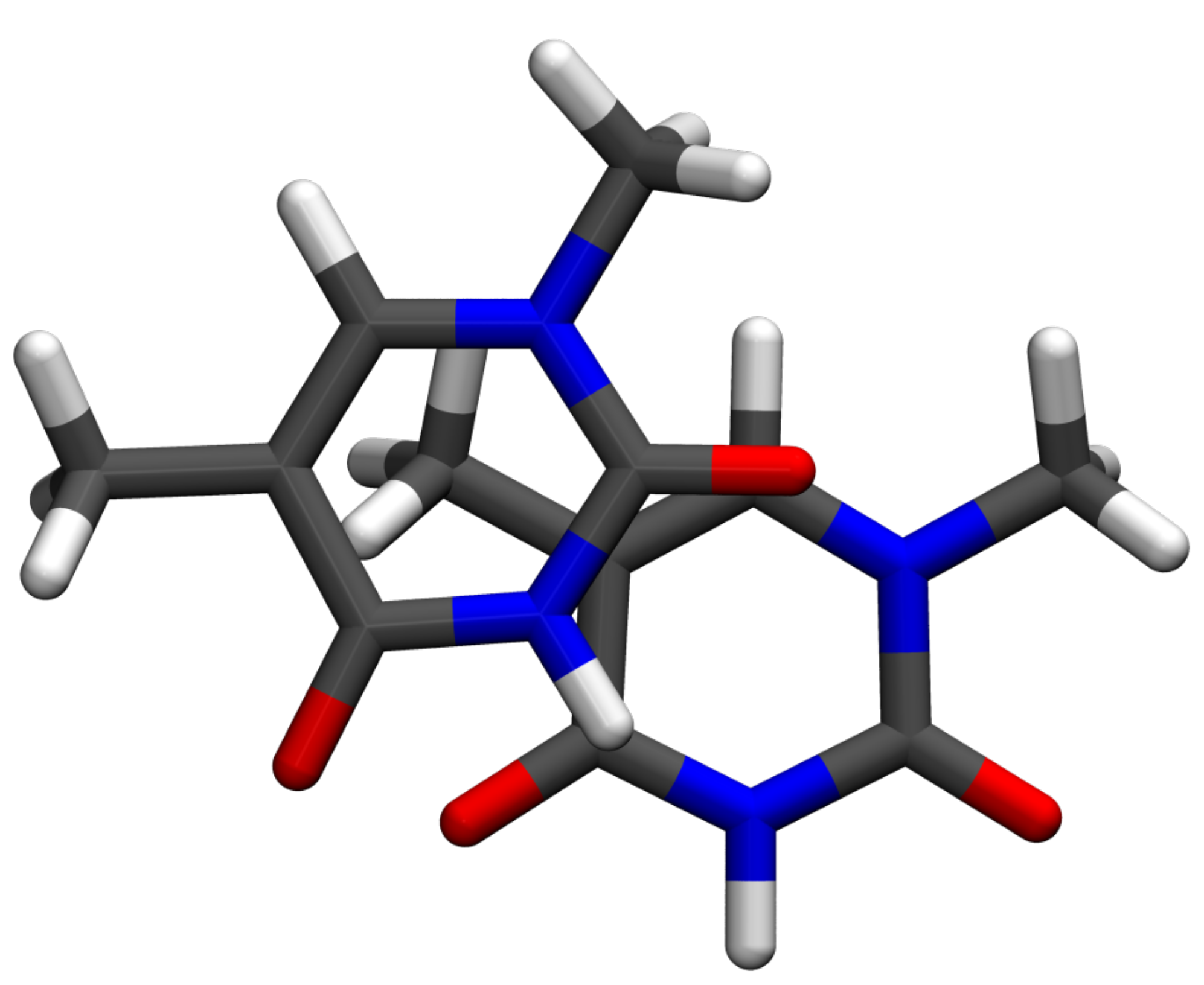}
        \end{center}
    \end{minipage}
\vline
    \begin{minipage}[c]{0.24\textwidth}     
        \begin{center}
            \includegraphics[width=\textwidth]{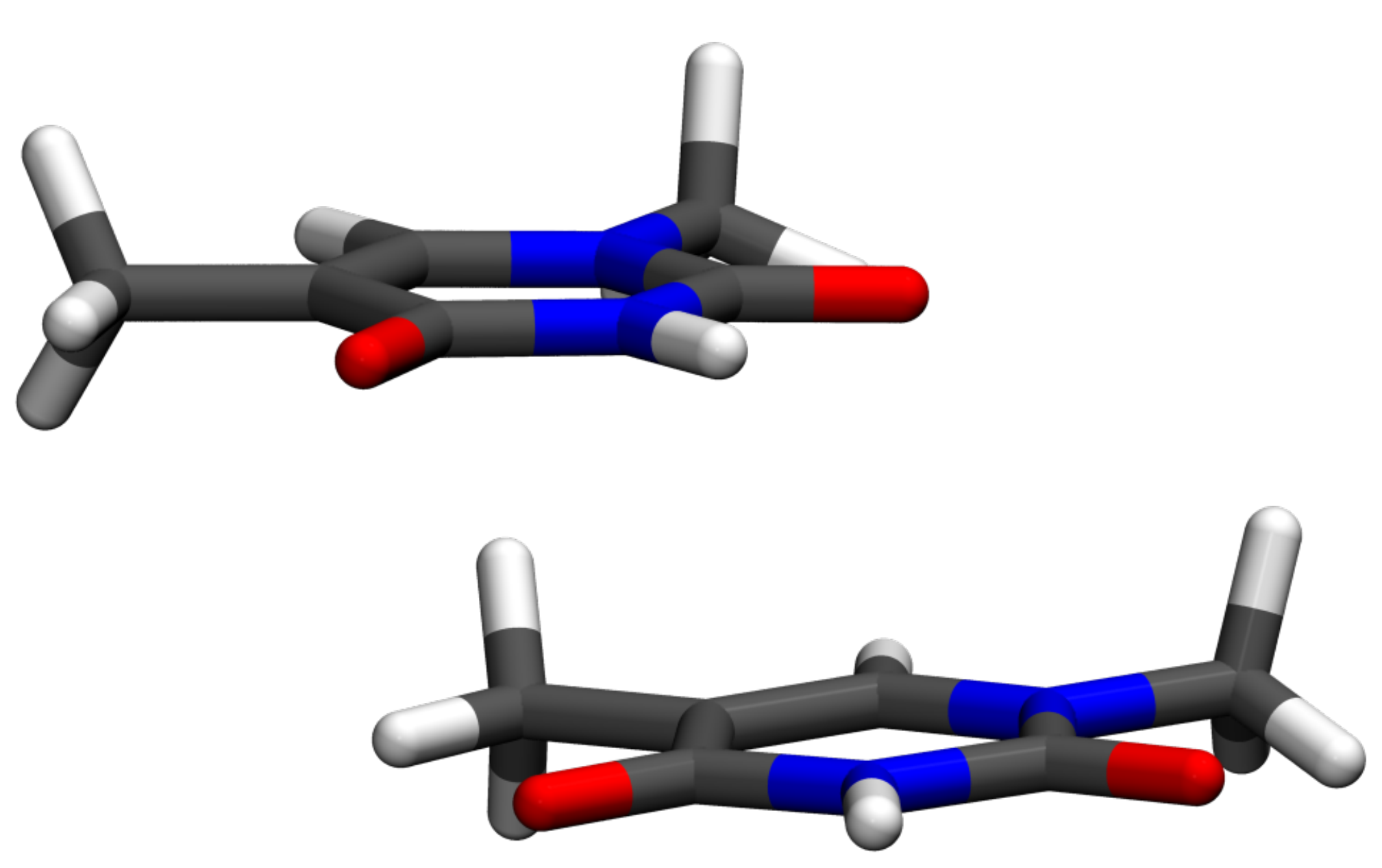}
        \end{center}
    \end{minipage}
    \begin{minipage}[c]{0.24\textwidth}     
        \begin{center}
            \includegraphics[width=\textwidth]{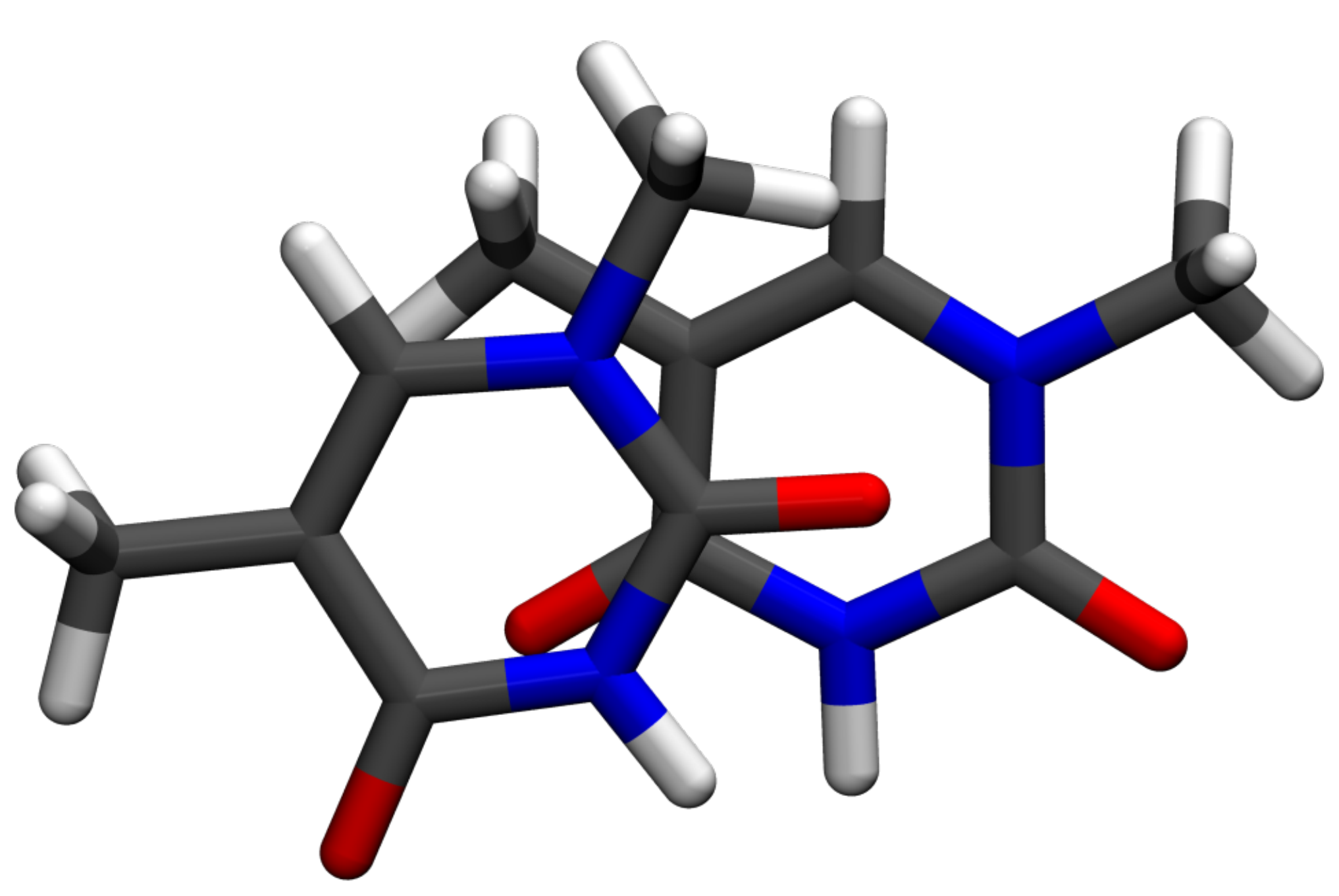}
        \end{center}
    \end{minipage}
\vspace{0.2cm}
    \begin{minipage}[c]{0.49\textwidth}     
        \begin{center}
            \footnotesize{A-DNA} 
        \end{center}
    \end{minipage}
    \begin{minipage}[c]{0.49\textwidth}     
        \begin{center}
            \footnotesize{B-DNA} 
        \end{center}
    \end{minipage}
\end{figure}
\centering
\begin{tabular}{l l l l l l l}
\hline
\hline
           & \multicolumn{6}{c}{XC Functional} \\
                     & LDA & BLYP & PBE & B3LYP & PBE0 & CAM-B3LYP \\
\hline
        \textbf{A-DNA} & & & & & & \\
        $\omega$  & 3.20229 & 3.18666 & 3.17333 & 3.29501 & 3.28174 & 3.33011 \\
              & 3.35610 & 3.30850 & 3.30172 & 3.40828 & 3.40349 & 3.45178 \\
\cline{2-7}
$\Delta\omega$ & 0.15381 & 0.12184 & 0.12839 & 0.11327 & 0.12175 & 0.12167 \\ 
\hline
        \textbf{B-DNA} & & & & & & \\
        $\omega$  & 3.17229 & 3.12597 & 3.12383 & 3.21472 & 3.21053 & 3.24955 \\
              & 3.24505 & 3.18026 & 3.17673 & 3.24457 & 3.23372 & 3.26865 \\
\cline{2-7}
$\Delta\omega$ & 0.07276 & 0.05429 & 0.05290 & 0.02985 & 0.02319 & 0.01910 \\
\hline
\hline
\end{tabular}
\label{tab:TT_DNABasePair_xcFunc}
\end{table}

\subsubsection*{b) PDI Dimer}
\begin{table}[H]
\caption{
Excitation energies $\omega$ (in eV) and splittings $\Delta\omega$ (in eV)
of the $\pi \rightarrow \pi^*$ triplet excitation of the PDI dimer 
for an inter-subsystem separation of 3.5~\r{A}
obtained from supersystem calculations
with different exchange--correlation functionals.
Two excitation energies are given because of the splitting of triplet excitations.
HOMO and LUMO of a PDI monomer are shown
[Supersystem: (LDA, BLYP, PBE, B3LYP, PBE0, CAM-B3LYP), def2-SVP].
}
\begin{figure}[H]
    \begin{minipage}[c]{0.49\textwidth}     
        \begin{center}
            \includegraphics[width=\textwidth]{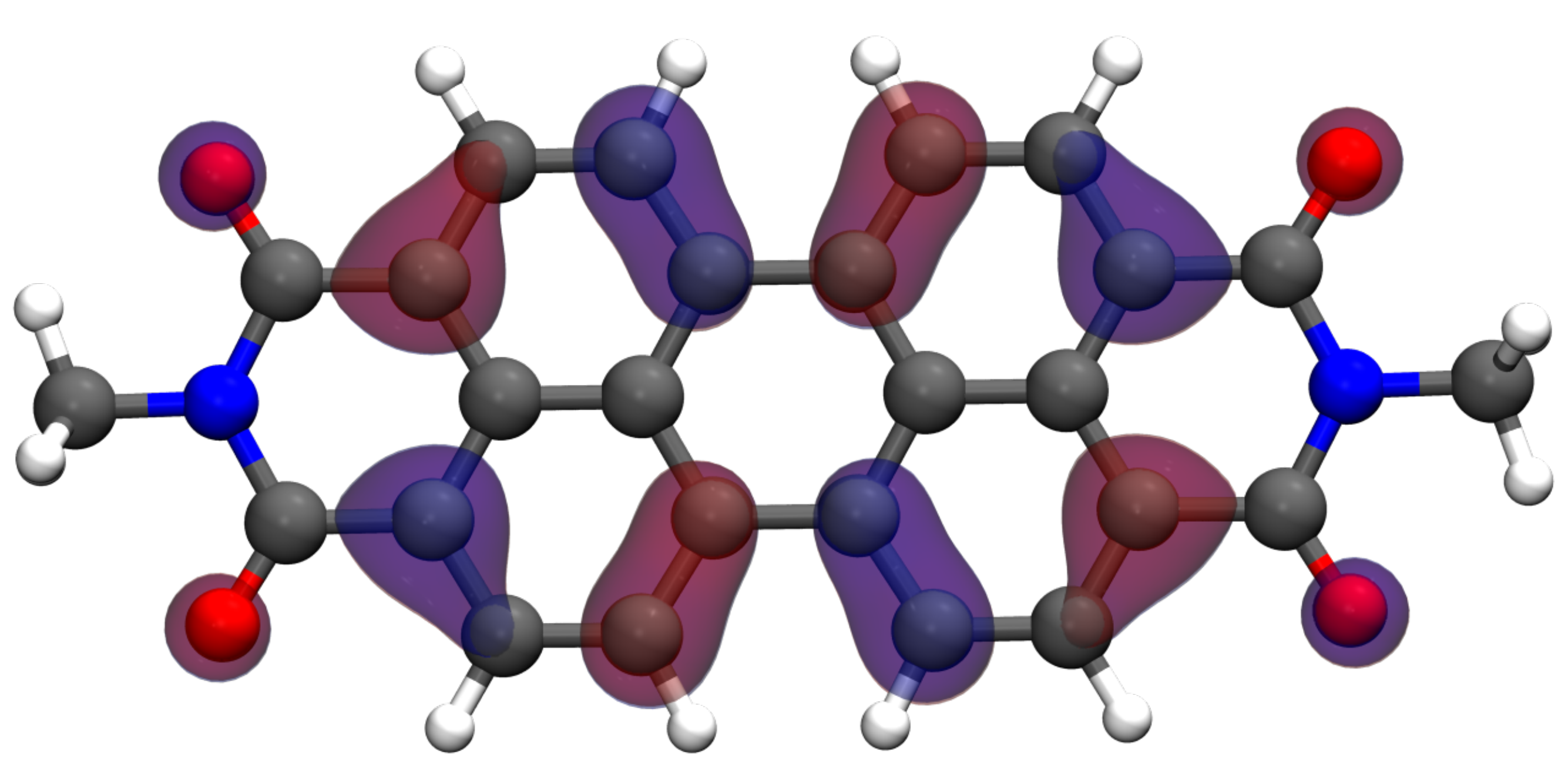}
        \end{center}
    \end{minipage}
\vline
    \begin{minipage}[c]{0.49\textwidth}     
        \begin{center}
            \includegraphics[width=\textwidth]{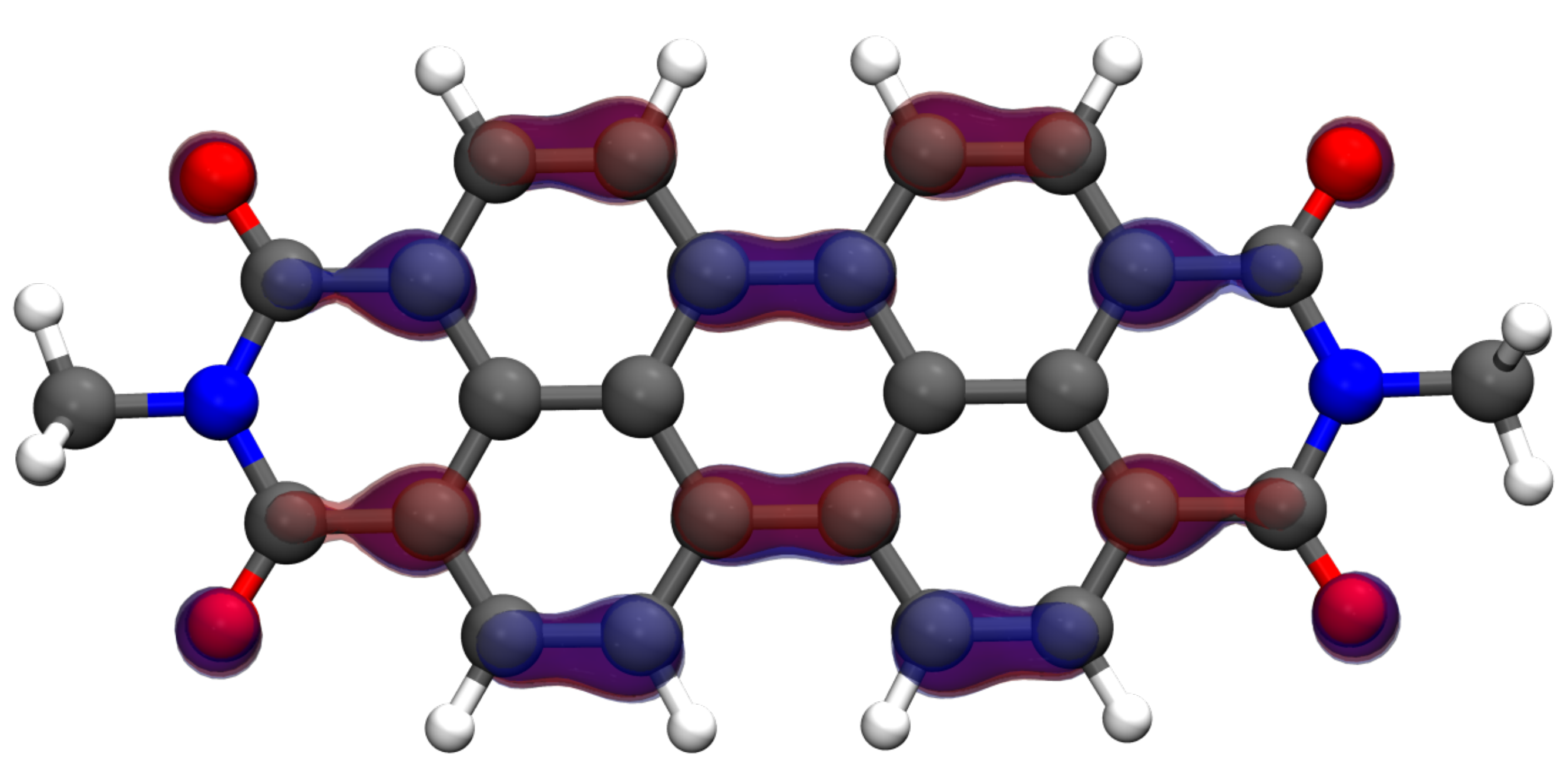}
        \end{center}
    \end{minipage}
\vspace{0.2cm}
    \begin{minipage}[c]{0.49\textwidth}     
        \begin{center}
            \footnotesize{HOMO} 
        \end{center}
    \end{minipage}
\hfill
    \begin{minipage}[c]{0.49\textwidth}     
        \begin{center}
            \footnotesize{LUMO} 
        \end{center}
    \end{minipage}
\vspace{0.3cm}
\end{figure}
\centering
\begin{tabular}{l l l l l l l}
\hline
\hline
                 & \multicolumn{6}{c}{XC Functional} \\
                 & LDA & BLYP & PBE & B3LYP & PBE0 & CAM-B3LYP \\
\hline
        $\omega$ & 0.87110 & 0.87167 & 0.87701 & 1.03143 & 1.05993 & 1.25130 \\
                 & 1.36873 & 1.32854 & 1.32680 & 1.41848 & 1.42304 & 1.54866 \\
\cline{2-7}
  $\Delta\omega$ & 0.49763 & 0.45687 & 0.44979 & 0.38705 & 0.36311 & 0.29736 \\
\hline
\hline
\end{tabular}
\label{tab:PDI_xcFunc}
\end{table}

\newpage
\section{Dependence of TEET Couplings on the Basis in NAKE-sTDDFT Calculations}
\begin{figure}[H]
    \begin{minipage}[c]{0.49\textwidth}     
        \begin{center}
            \includegraphics[width=\textwidth]{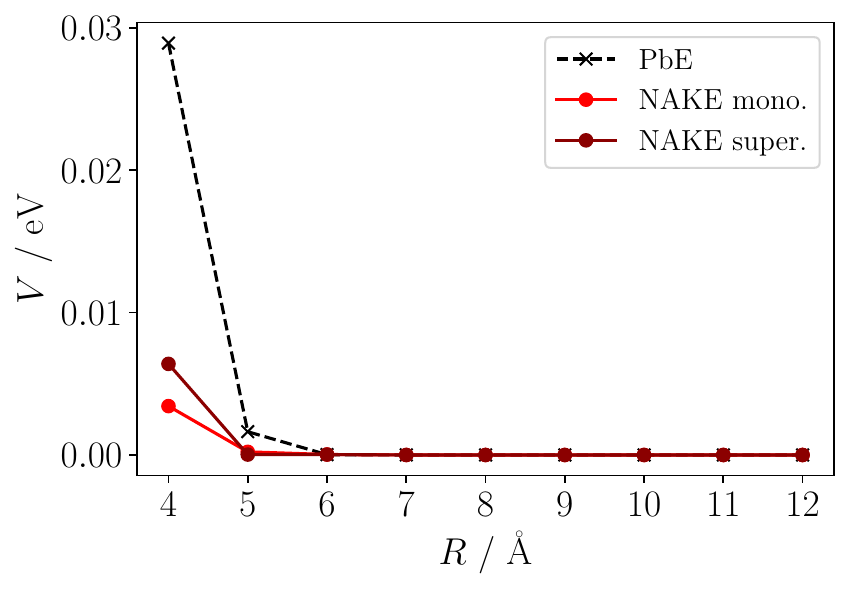} 
        \end{center}
    \end{minipage}
    \begin{minipage}[c]{0.49\textwidth}     
        \begin{center}
            \includegraphics[width=\textwidth]{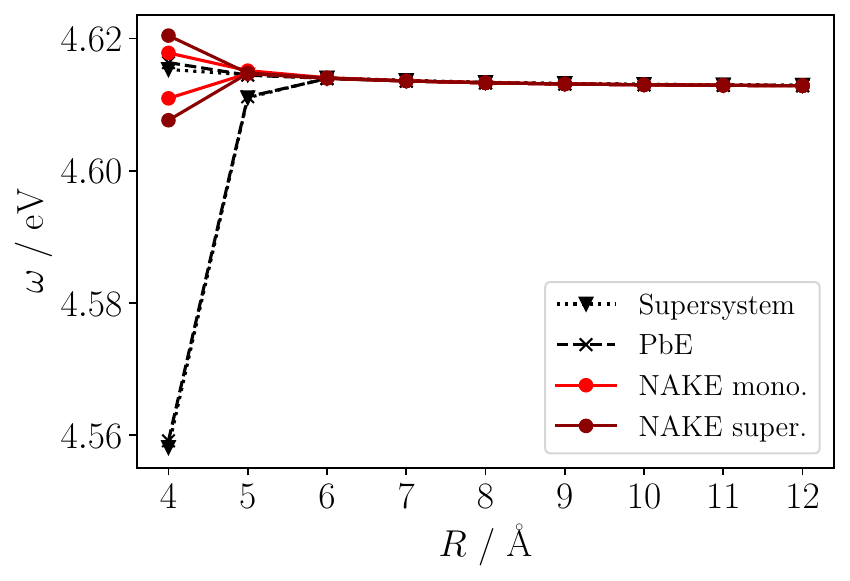} 
        \end{center}
    \end{minipage}
\caption{
Electronic couplings $V$ (left) and excitation energies $\omega$ (right) 
of the $\pi\rightarrow \pi^{*}$ triplet excitation as a function
of the inter-subsystem separation of the fluoroethylene dimer
[PbE: PBE0/PBE0/level.; NAKE: PBE0/PW91/PW91k; Supersystem: PBE0, def2-SVP].
}
\label{fig:PbE_vs_NAKE}
\end{figure}

\section{Influence of the Projection Operator} \label{HuzVsLevel}
As another aspect, the influence of the selected projection operator on the resulting
TEET couplings was explored. In addition to the levelshift operator applied in all 
PbE calculations presented above, we also test the Huzinaga operator in the following.
In contrast to the levelshift operator, the Huzinaga operator contains contributions
from the Fock matrix and may hence show a more intricate dependence on the 
inter-subsystem XC approximation.

\begin{figure}[H]
\captionsetup[subfigure]{justification=centering}
 \centering
    \begin{subfigure}[c]{0.32\textwidth}
        \includegraphics[width=\textwidth]{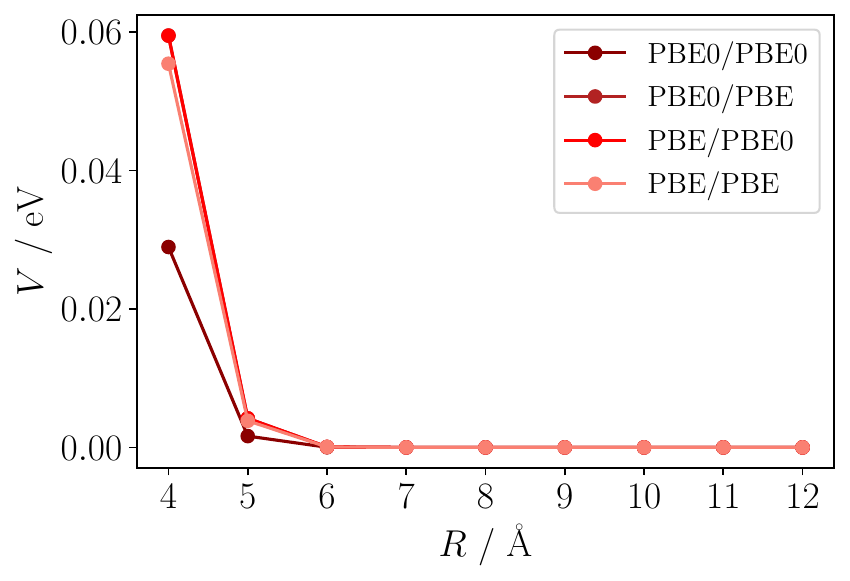}
        \includegraphics[width=\textwidth]{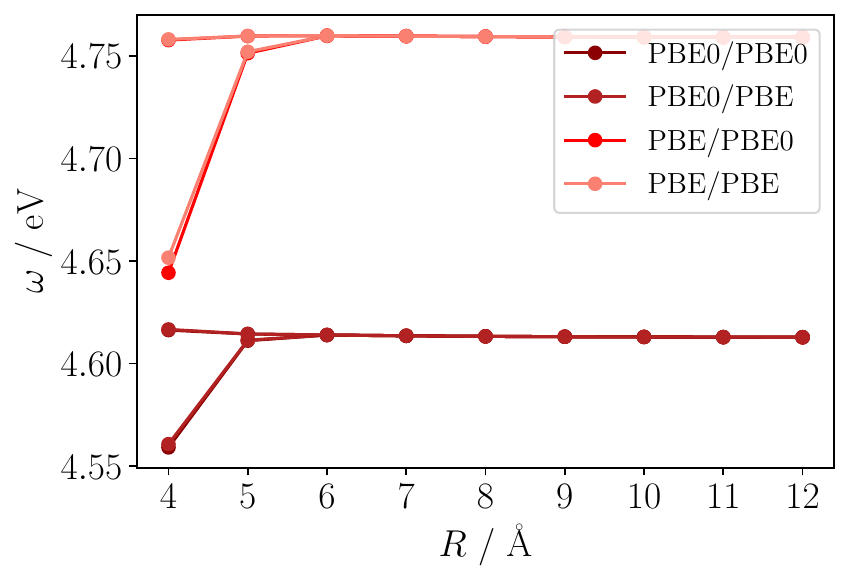}
        \caption{level.}
        \label{fig:Level}
    \end{subfigure}
    \begin{subfigure}[c]{0.32\textwidth}
        \includegraphics[width=\textwidth]{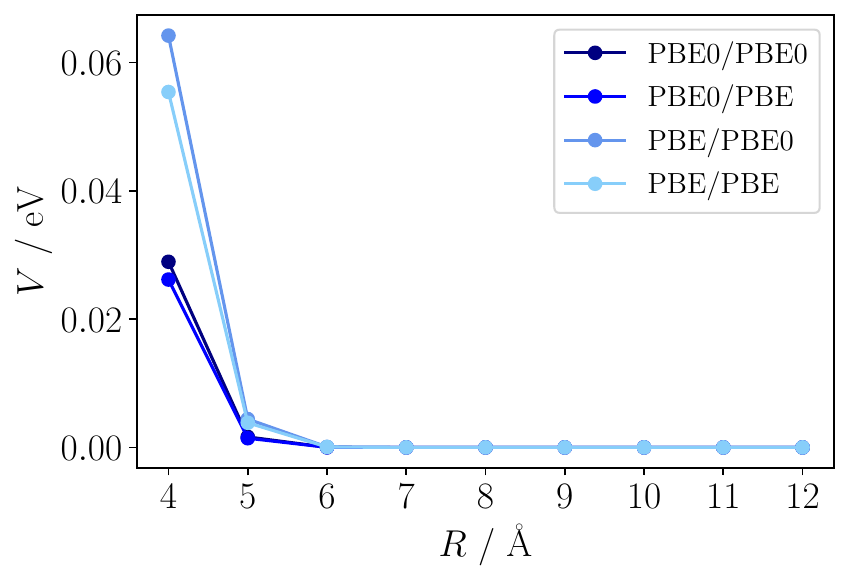}
        \includegraphics[width=\textwidth]{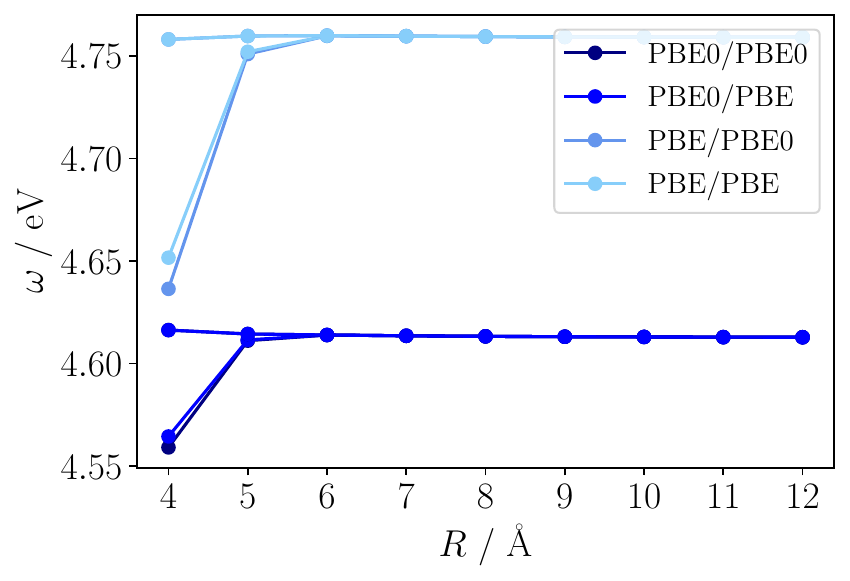}
        \caption{Huz.}
        \label{fig:Huz}
    \end{subfigure}
    \begin{subfigure}[c]{0.32\textwidth}
        \includegraphics[width=\textwidth]{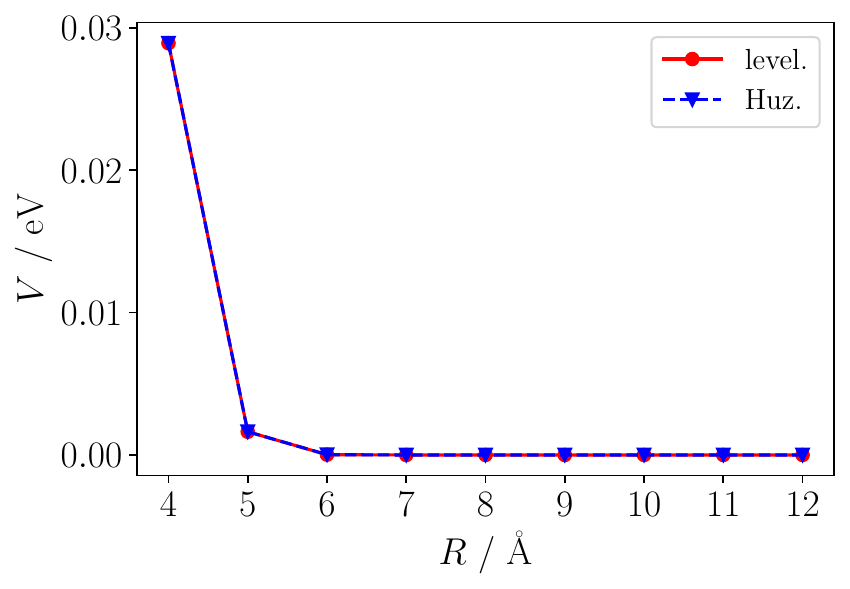}
        \includegraphics[width=\textwidth]{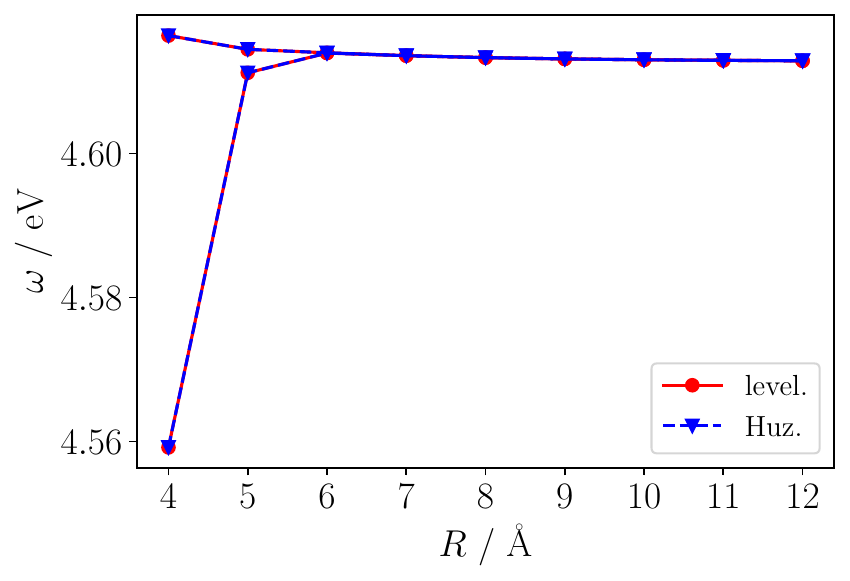}
        \caption{}
        \label{fig:LevelVsHuz}
    \end{subfigure}
\caption{
Electronic couplings (top) and excitation energies (bottom) 
of the  $\pi \rightarrow \pi^*$ triplet excitation 
as a function of the inter-subsystem separation 
of the fluoroethylene dimer 
obtained from PbE-sTDDFT using the \subref{fig:Level} levelshift 
and \subref{fig:Huz} Huzinaga projection operator
[PbE: (PBE0, PBE)/(PBE0, PBE)/(level., Huz.), def2-SVP], 
as well as \subref{fig:LevelVsHuz} a direct comparison of both projectors
[PbE: PBE0/PBE0/(level., Huz.), def2-SVP].
}    
\label{fig:Huz_Level}
\end{figure}

In Fig.~\ref{fig:Huz_Level}, electronic couplings and excitation energies for the 
fluoroethylene dimer system and using different combinations of the PBE0 and PBE
as the intra- and inter-subsystem XC functional and the def2-SVP basis set 
were determined employing either the levelshift or the Huzinaga operator. 
The results obtained are very similar for the two projection operators, 
which holds both for functionals with (PBE0) and without (PBE) exact exchange 
contributions.
Additional investigations were carried out concerning the dependence of the
Fock matrix contributions within the Huzinaga kernel.
Also for these contributions, it turns out that exact inter-subsystem exchange 
is not needed in order to obtain accurate results. 
This shows that there is no distinct advantage of the levelshift operator in terms 
of the dependence on the XC approximations.

\begin{table}[H]
\caption{
Impact of the levelshift parameter $\mu$ on the electronic coupling $V$
and excitation energies $\omega$ of the HOMO $\rightarrow$ LUMO excitation 
of the fluoroethylene dimer for an inter-subsystem separation of 4.0~\r{A}
obtained from PbE-sTDDFT calculations
[PbE: PBE0/PBE0/level., def2-SVP].
}
\centering
\begin{tabular}{c c c c}
\hline \hline
$\mu$ & $V$ / eV & $\omega_1$ / eV & $\omega_2$ / eV \\
\hline \hline
$10^2$ & 0.02893 & 4.55919 & 4.61637 \\
$10^3$ & 0.02893 & 4.55917 & 4.61637 \\
$10^4$ & 0.02893 & 4.55917 & 4.61637 \\
$10^5$ & 0.02893 & 4.55917 & 4.61637 \\
$10^6$ & 0.02893 & 4.55917 & 4.61637 \\
$10^7$ & 0.02893 & 4.55917 & 4.61637 \\
\hline \hline
\end{tabular}
\label{tab:my_label}
\end{table}

\section{Dependence of Bridge-Mediated TEET Couplings on the Number of Solvent Molecules} \label{PDI_SI}
\begin{figure}[H]
\captionsetup[subfigure]{justification=centering}
\centering
    \includegraphics[width=\textwidth]{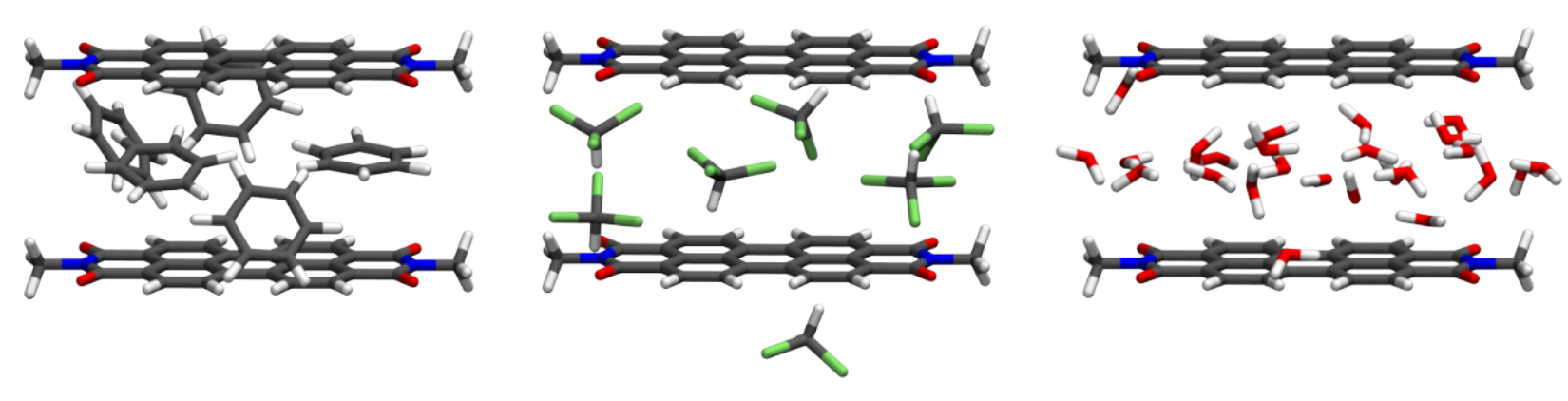}
    \begin{subfigure}[b]{0.33\textwidth}
        \includegraphics[width=\textwidth]{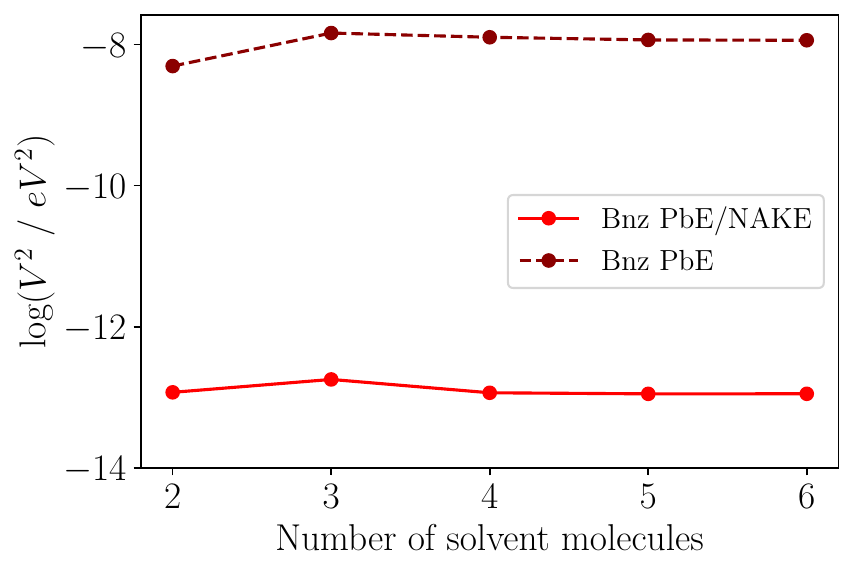}
        \caption{Benzene}
    \end{subfigure}
    \begin{subfigure}[b]{0.33\textwidth}
        \includegraphics[width=\textwidth]{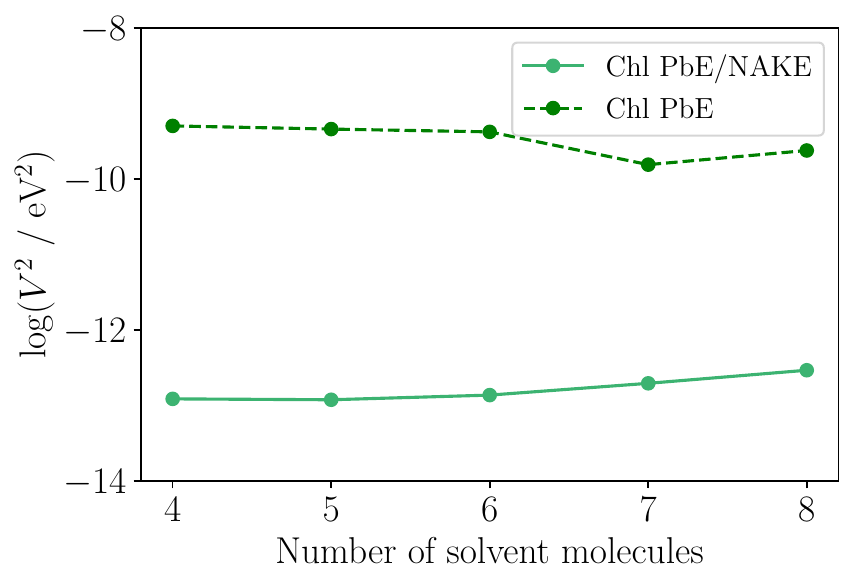}
        \caption{Chloroform}
    \end{subfigure}
    \begin{subfigure}[b]{0.32\textwidth}
        \includegraphics[width=\textwidth]{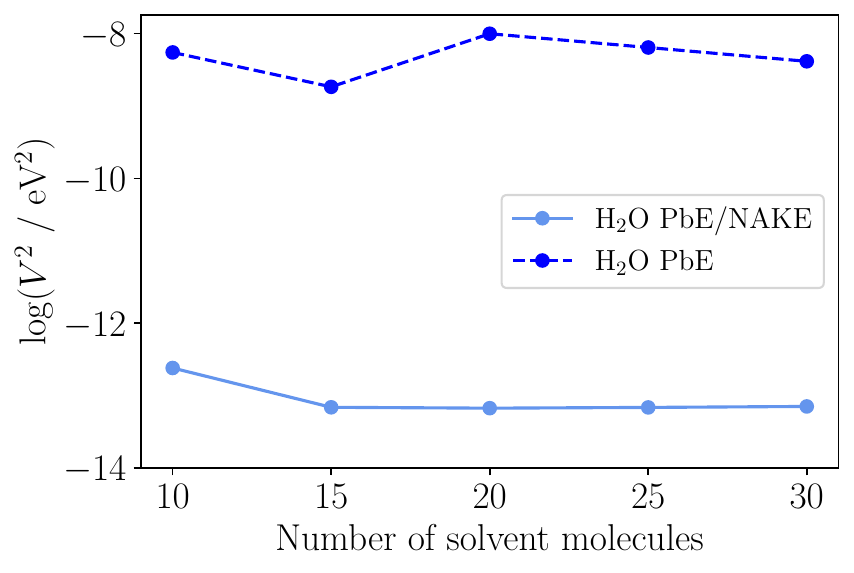}
        \caption{Water}
    \end{subfigure}
\caption{
Dependence of the logarithm of squared TEET couplings $V^2$ at $R = 7.0$~\r{A} 
inter-subsystem separations of the $\pi \rightarrow \pi^*$ triplet excitation
of the PDI dimer on the number of solvent molecules included in the model 
using water, chloroform (Chl), and benzene (Bnz)
calculated with mixed PbE-/NAKE-embedding and PbE-sTDDFT
[PbE: PBE0/PBE/level.;
PbE/NAKE: [PBE0, PBE]/[PBE, PBE]/[level., PW91k], def2-SVP].
}  
\label{fig:PDI_Curuchet_2}
\end{figure}

Based on structures provided by Carles Curutchet and used in Ref.~\cite{PDI.Curutchet.JPC.2012}, 
we systematically investigated the dependence of bridge-mediated couplings on the number of bridging 
solvent molecules, following a similar approach. For the sake of completeness, 
Fig.~\ref{fig:PDI_Curuchet_2} shows squared electronic 
couplings for increasing numbers of solvent molecules of one PDI dimer snapshot in 
different solvents ($R = 7.0$~\r{A}). Apparently, changing the number of solvent molecules
has only a minor effect on the TEET couplings for the various solvents once a certain 
minimal number of solvent molecules is included. 
Note that we employ the structures provided by Carles Curutchet and used in 
Ref.~\cite{PDI.Curutchet.JPC.2012} without further modifications. 
These calculations were only conducted for the purpose of comparing our method
for describing TEET couplings with the literature.

\section{Comparison of Bridge-Mediated TEET Couplings 
from the Fragment Excitation Difference Method and Subsystem TDDFT}
\begin{table}[H]
\caption{TEET couplings $V$ (in eV) 
at $R = 7.0$~\r{A} inter-subsystem separations 
of the $\pi \rightarrow \pi^*$ triplet excitations
of the PDI dimer with 25 water molecules located in between
calculated with PbE-sTDDFT [PBE0/PBE/level., def2-SVP] 
and the Fragment Excitation Difference (FED) method 
on the basis of TDA [PBE0/def2-SVP] and CIS calculations [HF/def2-SVP],
compared to data obtained in 
Ref.~\cite{PDI.Curutchet.JPC.2012} [HF/6-31+G(d)].
}
\centering
\begin{tabular}{c c c c c}
\hline
\hline
 System  &  PbE-sTDDFT  &  FED/TDA  &  FED/CIS  &  Ref. \\ 
\hline
\hline
 1  &  \num{8.01e-05}  &  \num{8.23e-05}  &  \num{1.32e-05}  &  \num{1.47e-05} \\
 2  &  \num{1.25e-04}  &  \num{1.31e-04}  &  \num{1.64e-05}  &  \num{1.65e-05} \\
 3  &  \num{8.30e-05}  &  \num{8.92e-05}  &  \num{1.02e-05}  &  \num{1.03e-05} \\
 4  &  \num{7.04e-05}  &  \num{7.14e-05}  &  \num{8.01e-06}  &  \num{9.40e-06} \\
 5  &  \num{1.51e-04}  &  \num{1.63e-04}  &  \num{1.97e-05}  &  \num{1.93e-05} \\
 6  &  \num{3.94e-05}  &  \num{3.96e-05}  &  \num{2.18e-06}  &  \num{3.04e-06} \\
 7  &  \num{1.46e-04}  &  \num{1.50e-04}  &  \num{1.37e-05}  &  \num{1.19e-05} \\
 8  &  \num{1.69e-05}  &  \num{1.77e-05}  &  \num{1.84e-06}  &  \num{1.53e-06} \\
 9  &  \num{3.43e-05}  &  \num{3.43e-05}  &  \num{3.38e-06}  &  \num{2.18e-06} \\
 10  &  \num{2.29e-04}  &  \num{2.39e-04}  &  \num{2.59e-05}  &  \num{2.54e-05} \\ 
\hline
\hline
\end{tabular}
\label{tab:FED_Ref}
\end{table}

\printbibliography
\clearpage